\documentclass[preprint]{aastex}
\received{ }
\accepted{ }

\slugcomment{Accepted for publication in {\it The Publications of the
Astronomical Society of the Pacific}}

\shorttitle{Spectropolarimetry of Type II Supernovae}
\shortauthors{Leonard \& Filippenko}

\newcommand{\kms}{km s$^{-1}$}
\newcommand{\Bband}{B}
\newcommand{\Vband}{V}
\newcommand{\BminusV}{({\Bband}{\rm -}{\Vband})}
\newcommand{\Ebv}{E\BminusV}
\newcommand{\ebv}{$E\BminusV$}
\newcommand{\halpha}{H$\alpha$}

\newcommand{\hbeta}{H$\beta$}

\newcommand{\vri}{\protect\hbox{$V\!RI$}}		
		
\newcommand{\ssp}{\def\baselinestretch{1.0}\large\normalsize}
\newcommand{\gtrsi}{\mathrel{\hbox{\rlap{\hbox{\lower4pt\hbox{$\sim$}}}\hbox{$>$}}}}

\begin{document}

\title{Spectropolarimetry of the Type II Supernovae 1997ds, 1998A, and 1999gi}

\vspace{2cm}

\author{Douglas C. Leonard\footnote{Present address: Department of Astronomy, 
University of Massachusetts, Amherst, MA 01003-9305}\ \ and Alexei V. Filippenko}
\affil{Department of Astronomy, University of California, Berkeley,
CA 94720-3411}
\email{leonard@nova.astro.umass.edu, alex@astro.berkeley.edu}

\vspace{1cm}

\begin{abstract}
We present single-epoch spectropolarimetry of the Type II supernovae (SNe II)
1997ds, 1998A, and 1999gi.  SN 1997ds and SN 1998A were both observed
during the early photospheric phase, less than 50 days after explosion, while
spectropolarimetry of SN 1999gi was obtained near the start of the transition
to the nebular phase, about 110 days after explosion.  Uncorrected for
interstellar polarization (ISP), SN 1997ds is characterized by $p_V = 0.85 \pm
0.02\%$, SN 1998A has $p_V = 0.24 \pm 0.05\%$, and SN~1999gi is polarized at
$p_V = 5.72 \pm 0.01\%$.  SN 1997ds and SN 1999gi exhibit distinct polarization
modulations (up to $\Delta p_{tot} = 1.6\%$ in SN~1997ds and $\Delta p_{tot} =
1.0\%$ in SN~1999gi) at the wavelengths of the strongest spectral line
features.  While no spectral polarization features could be observed in
SN~1998A, the data are insensitive to polarization features at the levels
confirmed in the other two objects.

The low continuum polarization inferred for SN~1997ds and SN~1998A and the
amplitude of (or limits on) the polarization modulations are consistent with those
measured at similar epochs for SN~1987A and the Type II-plateau SN~1999em, and
supports the growing consensus that core-collapse events with hydrogen
envelopes substantially intact at the time of explosion are not significantly
aspherical during the early photospheric phase.  The spectral shape of the high
continuum polarization of SN~1999gi closely resembles a ``Serkowski'' ISP curve
(characterized by $p_{\rm max} = 5.8\%, \theta = 154^\circ, {\rm and\ }
\lambda_{\rm max} = 5300$~\AA), and is inconsistent with the
wavelength-independent nature of electron scattering expected for an aspherical
SN atmosphere.  Since Galactic reddening is minimal along this line-of-sight,
the majority of the observed polarization in SN~1999gi is believed to be due to
ISP of the host galaxy, although significant (up to $p \approx 2\%$) intrinsic
polarization cannot be ruled out.  The potential power of SN
spectropolarimetry to study the properties of interstellar dust in external
galaxies is described and applied to the SN~1999gi data, where it is shown that
if the polarization is indeed predominantly interstellar in origin, then $R_V =
3.0 \pm 0.2$ for the dust along this line-of-sight in NGC~3184.

\end{abstract}

\medskip
\keywords {dust, extinction --- polarization --- supernovae: individual (SN
1997ds, SN~1998A, SN 1999gi) }

\section{INTRODUCTION}
\label{sec:introduction}

Polarimetry provides a powerful probe of early-time supernova (SN) geometry.
The fundamental idea is that a hot young SN atmosphere is dominated by electron
scattering, which by its nature is highly polarizing.  If we could resolve such
an atmosphere, we would measure changes in both the position angle and strength
of the polarization as a function of position in the atmosphere (see Figure~1
of Leonard, Filippenko, \& Matheson 2000b).  For a spherical source that is
unresolved, however, the directional components of the electric vectors cancel
exactly, yielding zero net linear polarization.  If the source is aspherical,
incomplete cancellation occurs; the resulting degree of net linear polarization
varies with the amount of asphericity, as well as with the viewing angle and
the extension and density of the electron-scattering atmosphere.  A
polarization of about 1\% is expected for an asymmetry of 30\% (defined as [1.0
- (minor axis / major axis)] $\times$ 100) for a typical viewing orientation
($45^\circ$ to the major axis) when modeled in terms of an oblate spheroid
(H\"{o}flich 1991).

From the Galactic distribution and high velocity of pulsars ($\sim 450$ \kms,
with individual velocities as high as 1600 km s$^{-1}$; see Cordes \& Chernoff
1998; Strom et al. 1995), high-velocity ``bullets'' of matter in SN remnants
(e.g., Taylor, Manchester, \& Lyne 1993), the asymmetric morphology of young SN
remnants (Manchester 1987; see, however, Gaensler 1998), the aspherical
distribution of material inferred by direct speckle imaging of young SNe (e.g.,
SN~1987A, Papaliolios et al. 1989; see, however, H\"{o}flich 1990), the
possible association of core-collapse SNe with some gamma-ray bursts (e.g.,
Bloom et al. 1999), and previous SN polarization studies (see Wheeler [2000]
for a comprehensive list of polarimetric observations of SNe), a consensus is
emerging that core-collapse SNe are intrinsically aspherical.  An interesting
recent speculation, however, is that the degree of asphericity is inversely
correlated with the amount of envelope material intact at the time of
explosion: SNe that have lost a significant amount of their hydrogen envelope
prior to exploding are found to be highly polarized at early times whereas
those that have their hydrogen envelopes substantially intact are polarized at
a much lower level (Wheeler 2000; see also Wang et al. 2001).  The observations
by Leonard et al. (2001a, hereafter L01a) that the polarization of SN~1999em, a
classic Type II event, slowly increased as the photosphere receded during the
plateau phase also supports the finding that the deeper we peer into the heart
of the explosion, the greater the evidence for asphericity becomes.  The number
of detailed spectropolarimetric studies, however, remains quite small, with
SN~1993J ($P \approx 1.5\%$, Tran et al. 1997) and SN~1998S ($P \approx 3\%$,
Leonard et al. 2000a\footnote{We note that the intense interaction of the SN
ejecta with a dense circumstellar medium in the SN~1998S system makes the
ultimate source of the inferred asymmetry (i.e., thermal photosphere or
ejecta-circumstellar medium interaction) unclear.}) serving as the prime
examples of SNe whose progenitors had shed a substantial fraction of their
hydrogen envelopes prior to explosion and SN~1987A ($P \approx 0.5\%$, Jeffery
1991a and references therein) and SN~1999em ($P \approx 0.3\%$, L01a) as the
standard bearers for those with substantially intact hydrogen envelopes.

We present single-epoch spectropolarimetry of SNe 1997ds, 1998A, and
1999gi, all Type~II events observed during the photospheric phase.  The
spectra are dominated by hydrogen Balmer lines with classic P-Cygni line
profiles, and there is no evidence for substantial circumstellar interaction or
progenitor mass-loss in any of the objects, suggesting the existence of
hydrogen envelopes largely intact at the times of the explosion.  We
describe the observations and reduction techniques in \S~\ref{sec:reductions},
the procedure used to set bounds on the interstellar polarization (ISP) and
upper detection limits for spectral and spectropolarimetric features in
\S~\ref{sec:ISP}, and the method used to quantify the strength of polarization
modulations in \S~\ref{sec:modulations}.  We present and discuss the
spectropolarimetry of the three SNe II in \S~\ref{sec:results}, and give an
overview of the potential power of using spectropolarimetry of SNe as probes of
interstellar dust in external galaxies in \S~\ref{sec:dustprobe}.  Our
conclusions are summarized in \S~\ref{sec:conclusions}.

\section{ANALYSIS TECHNIQUES}
\label{sec:analysistechniques}
\subsection{Data Reduction}
\label{sec:reductions}

All of the data presented in this study were obtained using the Low Resolution
Imaging Spectrometer (Oke et al. 1995) in polarimetry mode (Cohen 1996) at the
Cassegrain focus of the W. M. Keck-I and Keck-II 10-m telescopes, and reduced
according to the methods detailed by L01a.  All observations
were made with a 600 grooves/mm grating, blazed at 5000 \AA, which delivers a
resolution of $\sim 6$~\AA, with about 1.25~\AA\ pixel$^{-1}$.  A journal of
the spectropolarimetric observations is given in Table~1.  Measured values for
SNe~1997ds, 1998A, and 1999gi are listed in Table~2.

The displayed polarization $p$ is the ``rotated Stokes parameter'' (RSP),
calculated by rotating the normalized Stokes parameters $q$ and $u$ by a
smoothed fit to the polarization angle ($\theta$) so that all of the
polarization (in principle) falls in a single Stokes parameter (here, rotated
$q$).  To construct the fit to the polarization angle curve we first applied a
median filter to $\theta$ with a running boxcar of width 100~\AA, which helped
to eliminate extreme deviations in $\theta$ occurring over only one or two
pixels (as may result from cosmic-ray events, for example).  We then further
smoothed the resulting $\theta$ curve by replacing the value at each wavelength
with the mean of a 100~\AA\ boxcar centered on the wavelength.  Although
smoothing $\theta$ to construct the RSP in this way helps to preserve a
Gaussian distribution of polarization values about the true polarization level
when $\theta$ is a slowly varying function of wavelength (i.e., $\theta$ varies
on wavelength scales longer than the boxcar widths used to smooth it), sharp
polarization angle rotations will cause the computed RSP to underestimate the
true polarization level.  Since such sharp polarization angle rotations are
often seen in SN spectropolarimetry across line features, it becomes necessary to
examine both RSP and URSP (the polarization measured in the rotated $u$
parameter) to determine the true polarization of some line features.

\subsection{Setting Limits on Interstellar Polarization}
\label{sec:ISP}

A problem that plagues interpretation of all SN polarization measurements is
proper removal of the ISP: aspherical dust grains preferentially aligned with
the magnetic field of either the host galaxy or Milky Way (MW) can contribute a
polarization signal that dwarfs that of the SN.  For the three SNe~II studied
here, the most persuasive constraints that will be placed on the ISP come from
reddening estimates and the empirically derived relation
\begin{equation}
{\rm ISP}_{max} = 9\Ebv,
\label{eqn4:pmax}
\end{equation}
\noindent where ${\rm ISP}_{max}$ is the maximum ISP found for a given
reddening from observations of Galactic stars (Serkowski, Mathewson, \& Ford
1975; see also Whittet [1992, p. 83] for theoretical support for this
relation).  We estimate Galactic reddening from the dust maps of Schlegel,
Finkbeiner, \& Davis (1998).  To help characterize the ISP due to the MW
(ISP$_{MW}$), we also consider the polarization of MW stars near the
line-of-sight (l-o-s) to a SN contained in the catalog by Heiles (2000).  We
adopt the criterion of Tran (1995) that a MW star should be more than 150 pc
above (or below) the Galactic plane to fully sample the MW dust.  Although such
``probe'' stars will not, in general, yield the exact magnitude and direction
of the ISP$_{MW}$ along the true l-o-s (especially when the stars are more than
$\sim 1^\circ$ away), they should give a general indication of the polarization
in that region of the sky.

To estimate the reddening due to host-galaxy dust, we cautiously make use of the
rough correlation between the equivalent width ($W_\lambda$) of the \ion{Na}{1}
D interstellar lines and reddening found by Barbon et al. (1990),
\begin{equation}
\Ebv = 0.25 W_\lambda {\rm (Na\ I\ D)},
\label{eqn4:barbon}
\end{equation}
\noindent where $ W_\lambda {\rm (Na\ I\ D)}$ is the total equivalent width (in
\AA) of the interstellar \ion{Na}{1} D lines at $\lambda 5890 {\rm\ (D2)\ and\ }
\lambda 5896 {\rm\ (D1)}$.  We treat this relation with a healthy degree of
skepticism since sodium is known to be only a fair tracer of the hydrogen gas
column (especially in dense environments, where sodium may be heavily depleted;
e.g., Cohen 1973; cf., Phillips, Pettini, \& Gondhalekar 1984; Ferlet,
Vidal-Madjar, \& Gry 1985) which may then be used to estimate the dust column.
The dust-to-gas ratio also varies among galaxies (e.g., Issa, MacLaren, \&
Wolfendale 1990).  Perhaps the most important cause of uncertainty in assigning
a reddening value through this equation, though, is that our low-resolution
spectra do not resolve the individual absorption systems that contribute to the
line, making it impossible to estimate the effect of line saturation (i.e., the
point at which the curve-of-growth relating line width to number of gas atoms
begins to depart from a linear relation).  From analysis of echelle spectra of
32 O and early B stars suffering reddenings between $\Ebv = 0.06\ {\rm mag\
and\ } 1.56$ mag, Munari \& Zwitter (1997) find that saturation effects begin
to appear when $W_\lambda {\rm (Na\ I\ D)} \gtrsi 0.6$~\AA\ for a single
absorption system.  When the line becomes saturated, a measured equivalent
width may greatly underpredict the dust column and, hence, the reddening.  For
this reason, converting $W_\lambda {\rm (Na\ I\ D)}$ to an inferred reddening
value is considered to be most robust for low values of $W_\lambda$.

Barring high dust-to-gas ratios or unusually high depletion of Na gas along the
l-o-s, however, the lack of {\it any} \ion{Na}{1} D interstellar absorption
should, in principle, indicate little dust.  Since this may help place
constraints on the ISP, it is important to quantify the upper detection limit
of \ion{Na}{1} D absorption when no feature is apparent in a spectrum.  For
this, we adopt the relation of Hobbs (1984):
\begin{equation}
W_\lambda(3\sigma) = 3 \Delta\lambda\ \Delta I,
\label{eqn4:hobbs}
\end{equation}
\noindent where $W_\lambda(3\sigma)$ is the $3\sigma$ upper limit of the
equivalent width of a feature (in \AA), when $\Delta\lambda$ is the width of a
resolution element (in \AA) and $\Delta I$ is the 1$\sigma$ root-mean-square
fluctuation of the flux around a normalized continuum level.  This
equation is technically only correct for unresolved lines in a spectrum in
which the natural binning scale delivered by the spectrograph and CCD (i.e.,
the ``original binning'') provides 1 bin per resolution element.  A more
general equation, valid for both resolved and unresolved line features, is
\begin{equation}
W_\lambda(3\sigma) = 3 \Delta\lambda\ \Delta I \sqrt{W_{line} / \Delta\lambda\
} \sqrt{1 / B},
\label{eqn4:hobbs1}
\end{equation}
\noindent where $W_{line}$ is the width of the line feature (in \AA) and $B$ is the
number of bins per resolution element in a spectrum with the original binning.
For the data presented in this study, we take $B = 5$, and $W_{line} =
10$ \AA\ for the width of \ion{Na}{1} D interstellar absorption.

In a similar way, we shall also be interested in placing $3\sigma$ detection
limits on the strength of spectropolarimetric features in the normalized Stokes
parameters, $\Delta q (3\sigma)\ {\rm and\ }$ $\Delta u (3\sigma)$.  This is
more difficult to quantify than $W_\lambda (3\sigma)$ since we do not, in
general, know the intrinsic width of the expected features.  Indeed, some
polarization features in previous studies are only seen in one or two
(10~\AA\ bin$^{-1}$) pixels, while others stretch across a much wider range
(e.g., Leonard et al. 2000a; L01a).  For any individual pixel, the $3\sigma$
upper limit to detection is simply 3 times the $1\sigma$ uncertainty in the
normalized Stokes parameter for that pixel.  If an estimate is made of the
likely width of the feature, we shall derive an approximate upper detection
limit by dividing the $3\sigma$ upper limit for a single pixel by the number of
pixels examined.

\subsection{Quantifying Polarization Modulations}
\label{sec:modulations}

Since the ISP is not precisely known for the objects in this study, we
introduce a new definition to characterize the total polarization change seen
across a line profile that is independent of the ISP value:
\begin{equation}
\Delta p_{tot} \equiv \sqrt{\Delta q^2 + \Delta u^2},
\label{eqn4:1}
\end{equation}
\noindent where $\Delta q\ {\rm and\ } \Delta u$ are the individual changes in
the normalized Stokes parameters compared with the nearby continuum value
(i.e., $\Delta q \equiv q_{line} - q_{cont}$).  Note that $\Delta p_{tot}$ does
{\it not} measure the change in total {\it observed} polarization, $\Delta P$,
but is rather the square root of the quadrature sum of the individual changes
in the normalized Stokes parameters.  That is, a polarization change in which
the polarization vector only changes {\it direction} in the $q$-$u$ plane will
produce a measured polarization change of $\Delta P = 0$, but a non-zero,
positive value for $\Delta p_{tot}$.  In other words, $\Delta p_{tot}$ is a
measured quantity independent of the (generally unknown) ISP, whereas $\Delta
P$ is not.  Rather, $\Delta p_{tot}$ represents the maximum magnitude change in
the polarization, and makes no distinction between polarization increases,
decreases, or directional change.  Of course, since we do not know the exact
value of the ISP, interpretations dependent on whether the change is treated as
a polarization increase, decrease, or directional change necessarily do not
represent unique solutions.

When analyzing the polarization change across a specific line feature, we bin
the polarization spectrum to 10~\AA\ bin$^{-1}$ before determining the
polarization of a line feature in order to improve the signal-to-noise (S/N)
ratio.  However, since features may be quite sharp, it often happens that the
polarization change is detectable in only a few (rebinned) pixels.  In such
cases, we carefully examine the polarization properties of the original
spectrum (always binned 2~\AA\ bin$^{-1}$) to see if the polarization feature
is seen consistently across many pixels or is due to a single deviant pixel,
before concluding that a true polarization change exists.

In the spectropolarimetric study of SN~1999em by L01a, polarization increases
were found at the location of the P-Cygni line troughs of H$\beta$, \ion{Fe}{2}
$\lambda 5169$, and \ion{Na}{1} D $\lambda\lambda 5890, 5896$), and a sharp
polarization decrease was observed across the H$\alpha$ emission profile during
the photospheric phase.  These features were interpreted within the basic {\it
ansatz} first proposed by McCall (1984) and later extended by Jeffery
(1991b), that polarization peaks are naturally associated with absorption minima
due to selective blocking of forward-scattered (and hence less polarized) light
in P-Cygni absorption troughs and polarization minima are associated with
emission peaks due to the dilution of polarized continuum light with
unpolarized line emission.  The lack of a polarization increase in the
H$\alpha$ trough of SN~1999em was attributed to the increased contribution of
unpolarized photons by resonance scattering of continuum light along with the
fact that the optically thick \halpha\ line provides a screen that effectively
covers more of the SN atmosphere than the metal lines, thereby blocking
relatively more polarized photons from the limb regions (see discussion by
L01a).

With this simple model in mind, it is possible to solve for a lower bound on
the true continuum polarization level ($p_{cont}$) in the limit of blockage of
only unpolarized light (or light with electric vectors that exactly cancel) in
a line trough and addition of completely unpolarized light in a line peak:
\begin{equation}
p_{cont} \geq \frac{\Delta p_{tot}} {(I_{cont}/I_{trough}) - 1},
\label{eqn4:pcont}
\end{equation}
\noindent where $I_{cont}$ equals the interpolated value of the continuum flux
at the location of the line feature and $I_{trough}$ is the total flux at the
line's flux minimum.  Equation~(\ref{eqn4:pcont}) will allow us to set
plausible bounds on the continuum polarization intrinsic to a SN, and not
produced by ISP.  To estimate the continuum
flux level at the location of an absorption feature, we fit a smooth spline
over the absorption feature, guided by the continuum levels shown for the
synthetic spectra of SNe~II presented by Jeffery \& Branch (1990).  

L01a presents a detailed spectropolarimetric study of SN~1999em, an extremely
well-observed classic Type-II plateau (SN II-P) event, and Leonard et
al. (2001b, hereafter L01b) present and discuss its spectral and photometric behavior.
Since the total flux spectra of SNe~1997ds, 1998A, and 1999gi are all
broadly similar to SN~1999em at similar epochs, direct comparisons between
these three events and SN~1999em will frequently be made.

\addcontentsline{lot}{table}{\protect\numberline{1}Journal of
Spectropolarimetric Observations of SN~1997ds, SN~1998A, and SN~1999gi}
\begin{figure}
\vskip 0.2in
\hskip -2.5in
%\begin{center}
\rotatebox{90}{
\scalebox{1.1}{
\plotone{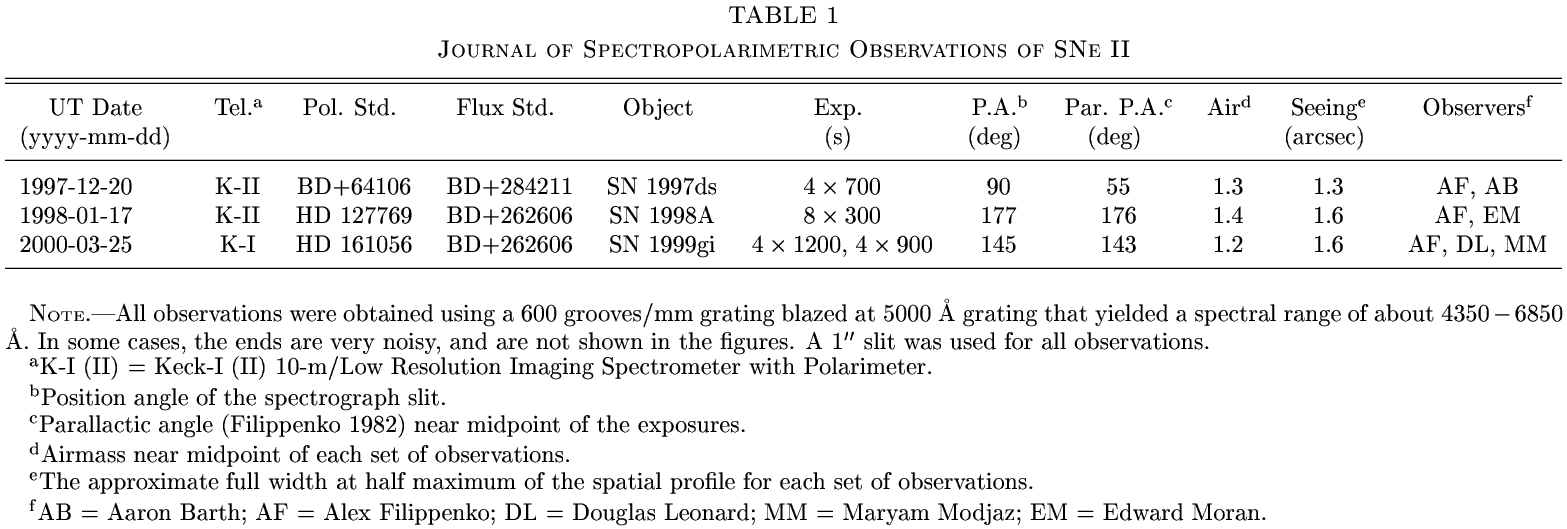}
} }
%\end{center}
\end{figure}

\addcontentsline{lot}{table}{\protect\numberline{2}Spectropolarimetry Data
for SN~1997ds, SN~1998A, and SN1999gi}
\begin{figure}
\vskip 0.2in
\hskip -2.6in
%\begin{center}
\rotatebox{90}{
\scalebox{1.1}{
\plotone{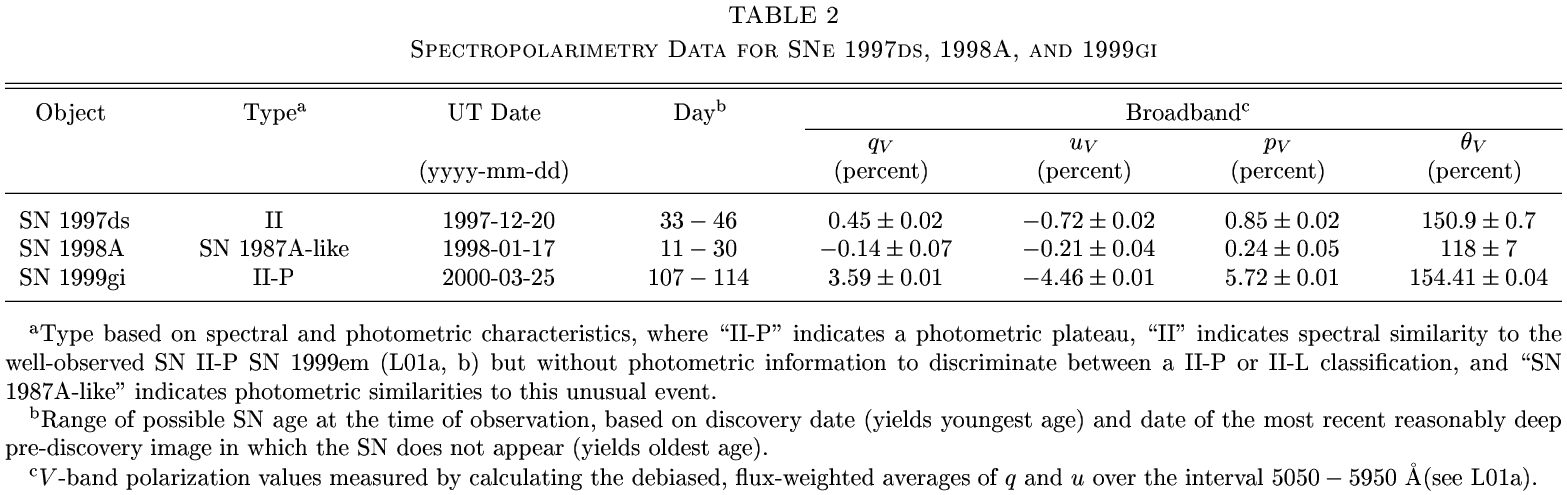}
} }
%\end{center}
\end{figure}

\section{Results and Discussion}
\label{sec:results}
\subsection{SN~1997ds}
\label{sec:sn1997ds}

SN~1997ds was discovered by Qiu et al. (1997) on 1997 November 17.47 at
unfiltered magnitude $m \approx 16.2$ mag in the SBd galaxy MCG -01-57-007.
Patat, Boehnhard, \& Delfosse (1997) obtained a spectrum 3 days later and
identified it as a Type~II event, with well-developed \halpha\ and \hbeta\
P-Cygni profiles.  Since a CCD image of the same field taken 13 days earlier
did not detect the SN (Qiu et al. 1997), it was likely discovered quite soon
after explosion.  The Galactic foreground extinction is small, $\Ebv_{MW} =
0.082$ mag (Schlegel et al. 1998), and the lack of appreciable interstellar
\ion{Na}{1} D absorption ($W_\lambda [3\sigma] < 0.03$~\AA, from 
equation~[\ref{eqn4:hobbs1}]) suggests minimal host-galaxy reddening as well.

\begin{figure}
\ssp
\begin{center}
\rotatebox{0}{
 \scalebox{0.8}{
	\plotone{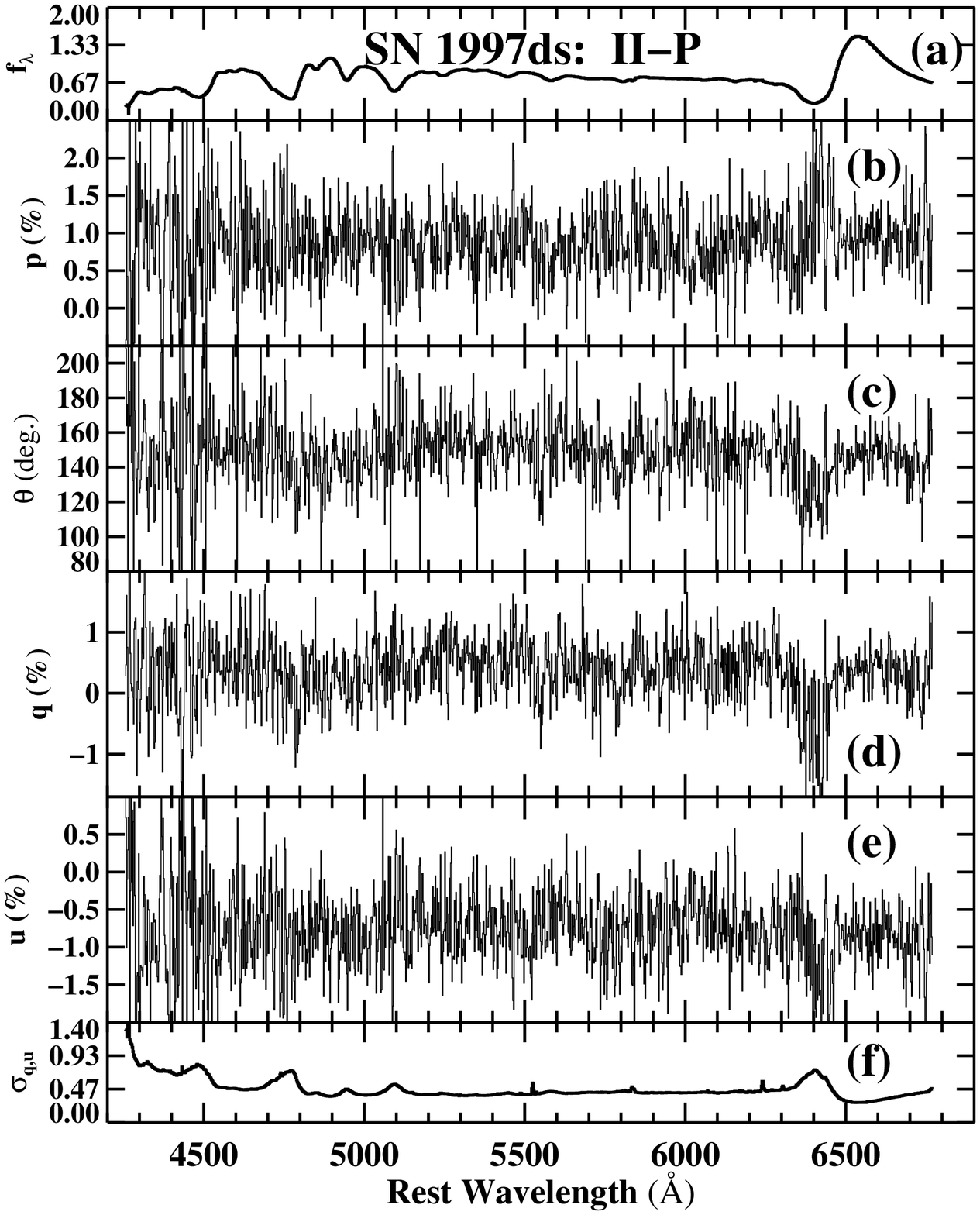}
		}
		}
\end{center}
\caption[Polarization data for SN~1997ds]
{Polarization data for SN~1997ds obtained 1997 December 20, between 33 and 46
days after explosion.  The NASA/IPAC Extragalactic Database (NED) recession
velocity of 2833 km s$^{-1}$ for MCG -01-57-007 has been removed in this and
all figures.  ({\it a}) Total flux, in units of $10^{-15}$ ergs s$^{-1}$
cm$^{-2}$
\AA$^{-1}$.  ({\it b}) Observed degree of polarization.  ({\it c}) Polarization
angle in the plane of the sky. ({\it d, e}) The normalized $q$ and $u$ Stokes
parameters. ({\it f}) Average of the (nearly identical) $1\sigma$ statistical
uncertainties in the Stokes $q$ and $u$ parameters for the displayed binning of
2~\AA\ bin$^{-1}$.  The polarization shown in this and all plots is the
``rotated Stokes parameter''; see text for details.
}
\label{fig4:1_97ds}
\end{figure}
%\clearpage

\begin{figure}
\ssp
\vskip -0.5in
\hskip -0.3in
%\begin{center}
\rotatebox{90}{
 \scalebox{0.8}{
	\plotone{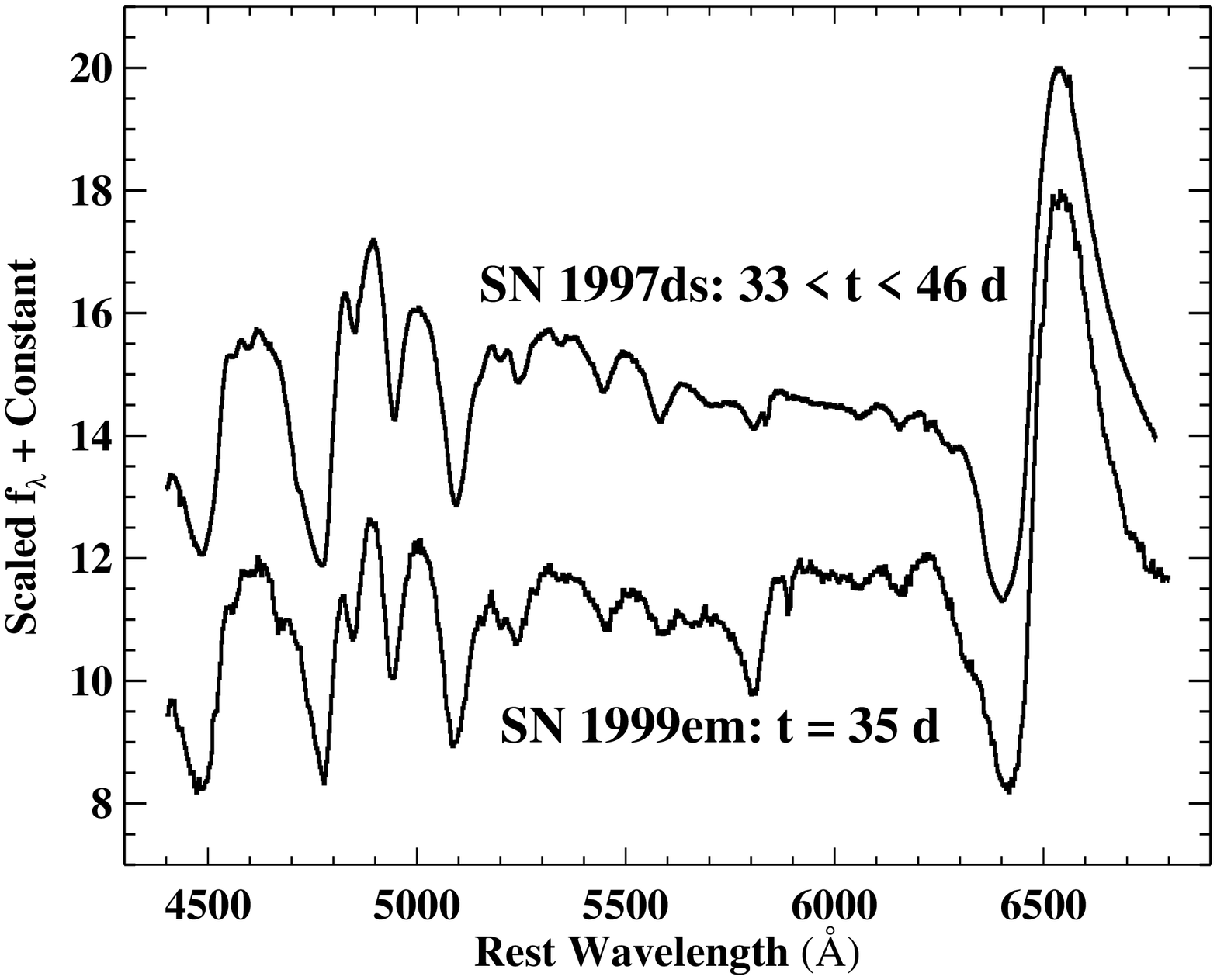}
		}
		}
%\end{center}
\caption[Total flux spectra of SN~1997ds and SN~1999em compared] {The total
flux spectrum of SN~1997ds ({\it top}) and SN~1999em ({\it bottom}), with the
estimated time since explosion indicated.  The spectra have been offset from
each other for clarity.  In this and all figures the explosion date of
SN~1999em is taken to be 5 days before discovery (L01b).  The close resemblance
of all the features suggests a comparable evolutionary phase for the two events.  A
redshift of 800 \kms\ has been removed from the spectrum of SN~1999em in this
and all figures.}
\label{fig4:5_97ds}
\end{figure}

We obtained spectropolarimetry of SN~1997ds on 1997 December 20 at the Keck-II
10-m telescope (Table~1); some preliminary results from this observation were
presented by Leonard et al. (2000b). From the constraints set
by the discovery and prediscovery images, the observation occurred sometime
between $33$ and $46$ days after the explosion, during the early
photospheric epoch.  The data are presented in Figure~\ref{fig4:1_97ds} and
tabulated in Table~2. The total flux spectrum shows the usual features for a
SN~II during the photospheric phase: prominent hydrogen Balmer and \ion{Fe}{2}
P-Cygni features superposed on a fairly smooth continuum.  Although no
photometric information is available for SN~1997ds, we note that the total flux
spectrum closely resembles SN~1999em at a similar age
(Figure~\ref{fig4:5_97ds}), which might suggest that SN~1997ds belongs to the
Type II-P classification as well.  However, the lack of sufficient
spectroscopic data for Type~II linear (II-L) SNe in general precludes
definitive classification based on spectroscopy alone (Patat et al. 1994; cf.,
Schlegel 1996).

The spectropolarimetry data displayed in Figure~\ref{fig4:1_97ds}(b,c) indicate
a low, flat polarization at $p \approx 0.85\%$ with a fairly constant
polarization angle near $\theta = 151^\circ$.  From the foreground reddening,
we infer a maximum ISP due to MW dust of ${\rm ISP}_{MW} < 0.74\% $
(equation~[\ref{eqn4:pmax}]), and an unknown, though probably minimal
contribution from the host galaxy (${\rm ISP}_{host} < 0.07\% $, if we apply
equation~[\ref{eqn4:barbon}] to the detection limit of the \ion{Na}{1} D
interstellar lines).  Further support for a small Galactic ISP comes from the
polarization of stars near the l-o-s to SN~1997ds.  The cataloged values
(Heiles 2000) for the polarization of 8 stars within $5^\circ$ of SN~1997ds are
quite small, ranging from $p = 0.003 \pm 0.001\% $ for HD 213998 to $p = 0.869
\pm 0.061\% $ for HD 212571.  The polarization angles given for these 8 stars
are essentially random, although only 2 (HD 212571 and HD 211099 [$p = 0.26 \pm
0.038\% $]) lie sufficiently far away (i.e., more than 203 pc for the l-o-s to
SN~1997ds, which has a Galactic latitude of $-48^\circ$) to fully probe the
material in the Galactic plane.  Given the range of values found for the
foreground stars, it is possible that a significant portion of the observed
polarization of SN 1997ds is due to Galactic dust.  Lacking temporal coverage
or a clear physical motivation from which to deduce the ISP value, however, we
cannot be certain.  We do note that if the maximum allowed host ISP value is
used, then an upper limit to the total ISP (i.e., ISP$_{MW}$ + ISP$_{host}$) of
ISP$_{tot} < 0.81\%$ results.  This then restricts the allowable polarization
intrinsic to SN~1997ds to the range $0.04\% \leq p \leq 1.66\%$.

\begin{figure}
\ssp
\begin{center}
\rotatebox{0}{
 \scalebox{0.8}{ \plotone{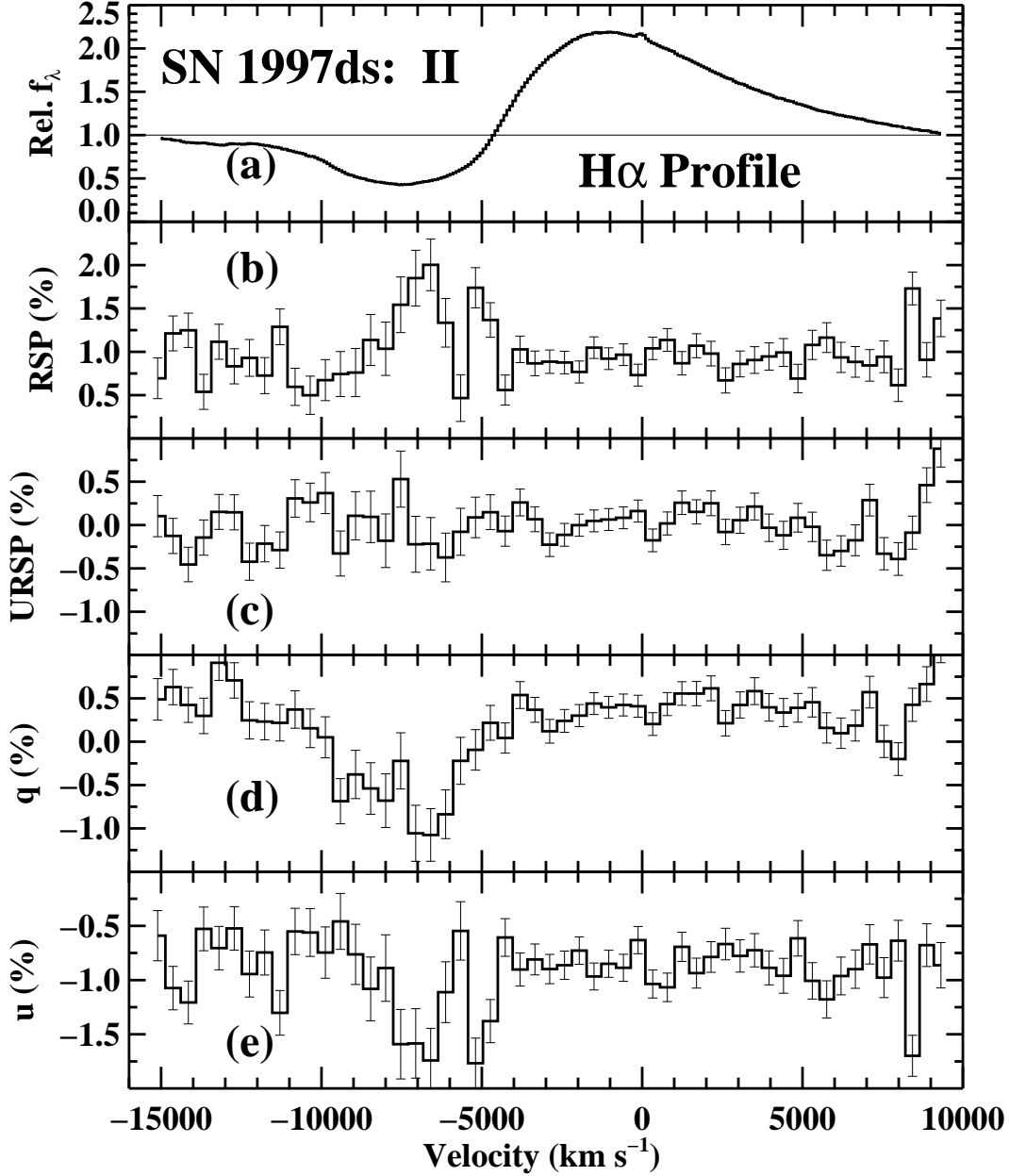}
	} }
\end{center}
\caption[Spectropolarimetry of H$\alpha$ region for SN~1997ds] {Region
near H$\alpha$ for SN~1997ds on 1997 December 20, between 33 and 46 days
after the explosion.   Error bars are $1\sigma$ statistical for
10~\AA\ bin$^{-1}$.  ({\it a}) Total flux, normalized by a spline fit to the
continuum, displayed at 2~\AA\ bin$^{-1}$ for better resolution.  ({\it b})
Observed degree of polarization, determined by the rotated Stokes parameter
(rotated $q$).  ({\it c}) Observed degree of polarization in the rotated $u$
parameter.  ({\it d, e}) Normalized $q$ and $u$ Stokes parameters.}
\label{fig4:2_97ds}
\end{figure}

A clear line feature in the polarization data is visible at \halpha\
(Figure~\ref{fig4:1_97ds}); Figure~\ref{fig4:2_97ds} shows more detail of this
region, rebinned to 10~\AA\ bin$^{-1}$ to improve the S/N ratio.  A change of
$\Delta q \approx -1.3\%$ and $\Delta u \approx -0.7\%$ is measured in the
deepest part of the P-Cygni trough, resulting in a total polarization change of
$\Delta p_{tot} \approx 1.6\% $ (equation~[\ref{eqn4:1}]) for the line.  (We
note that the single-pixel, $3\sigma$ upper limit for polarization changes at
the \hbeta\ trough is $\Delta p_{tot} [3\sigma] = 1.36\% $, and at \ion{Fe}{2}
$\lambda 5169$ is $\Delta p_{tot} [3\sigma] = 1.02\%$.)  Since we do not know
the ISP, it is impossible to state with certainty whether the intrinsic
polarization increases, decreases, or changes direction in this feature.  In
the spectropolarimetry of SN~1999em, a sharp polarization drop occurs across
H$\alpha$ from just redward of the absorption minimum through the emission
profile (L01a).  For SN~1997ds, the polarization modulation seen at H$\alpha$
is distinctly different from that seen in SN~1999em: independent of the ISP
value, the strongest part of the modulation is clearly associated with the
absorption trough itself, with little modulation evident across the emission
profile.  Further, if the ISP contribution is less than 0.81\% as expected,
then the polarization in the trough {\it cannot} represent a polarization
decrease.  An \halpha\ trough polarization increase is actually quite similar
to what is observed in the \ion{Fe}{2} $\lambda 5169$ and \ion{Na}{1} D
$\lambda\lambda 5890, 5896$ lines of SN~1999em (L01a), suggesting that the gas
forming the \halpha\ absorption line in SN~1997ds screens relatively fewer
polarized photons than it does in SN~1999em.  The different characteristics of
the \halpha-line polarization between SN~1997ds and SN~1999em may signal
a different density structure of the electron-scattering atmospheres of the two
events. 

Lacking a firm value for the ISP, however, it is difficult to proceed with
further interpretation of the \halpha\ polarization feature.  The basic result
is that at the time of our observation, between 33 and 46 days after the
explosion, SN~1997ds had a rather low intrinsic polarization, most likely in
the range $0.04\% \leq p \leq 1.66\%$.  A distinct polarization modulation of
$\Delta p_{tot}
\approx 1.6\% $ occurs in the \halpha\ absorption trough, and polarization modulations
greater than $\sim 1.3\%$ in other absorption features are not seen.

\subsection{SN~1998A}
\label{sec:sn1998a}

SN~1998A was discovered by Williams et al. (1998) on 1998 January 6.77 at an
$R$-band magnitude of $m_R \approx 17$ mag in the Sc galaxy IC~2627.  Since the
SN is not visible on a CCD image of the same field taken 20 days earlier
(Williams et al. 1998), it was likely discovered shortly after the explosion.
Filippenko \& Moran (1998) identified it as a Type~II event with well-developed
P-Cygni hydrogen Balmer lines from a spectrum taken on 1998 January 17 (and
presented here).  The foreground reddening to SN~1998A is minimal, \ebv $_{MW}
= 0.07$ mag (Schlegel et al. 1998), and the lack of appreciable interstellar
\ion{Na}{1} D absorption ($W_\lambda [3\sigma] < 0.13$~\AA\ ) suggests small
host galaxy reddening as well, with $\Ebv_{host} < 0.03$ mag from
equation~(\ref{eqn4:barbon}).  This implies $\Ebv_{tot} < 0.1$ mag and ${\rm
ISP}_{max} = 0.9\% $.  The polarization catalog by Heiles (2000) includes
measurements for 8 MW stars within $5^\circ$ of SN~1998A, seven of which have
$p < 0.3\%$, with one (HD 94473) having $p = 1.0 \pm 0.03\%$.  Several of these
stars are far enough away (i.e., more than 255 pc since SN~1998A has a Galactic
latitude of $34^\circ$) to fully probe the material in the Galactic plane.  All
8 of the MW stars have a similar polarization angle ($49^\circ <
\theta < 89^\circ$), which suggests fairly ordered but patchy dust in this
direction (see Ryu et al. 2000 for a detailed study of the patchy extinction in
this region).

A photometric study of SN~1998A during the first 170 days after discovery is
presented by Woodings et al. (1998).  The \vri\ light curves show a slow rise
to a broad ($\sim 30$ days) peak approximately 70 days after discovery,
followed by a more rapid decline.  This light curve behavior is quite similar
to that of SN~1987A, leading Woodings et al. (1998) to propose that the
progenitor of SN~1998A may also have been a blue supergiant (Woosley et
al. 1987).  If this comparison is valid, then we might expect SN~1998A to share
some of the spectral properties that were seen in SN~1987A as well, the most
prominent of which was the very rapid strengthening of the metal lines (in
particular lines resulting from \ion{Ba}{2}; see, e.g., Jeffery \& Branch 1990)
during the early photospheric phase, an effect believed to result from the
compact nature of its progenitor (Branch 1987).

\begin{figure}
\ssp
\begin{center}
\rotatebox{0}{
 \scalebox{0.8}{
	\plotone{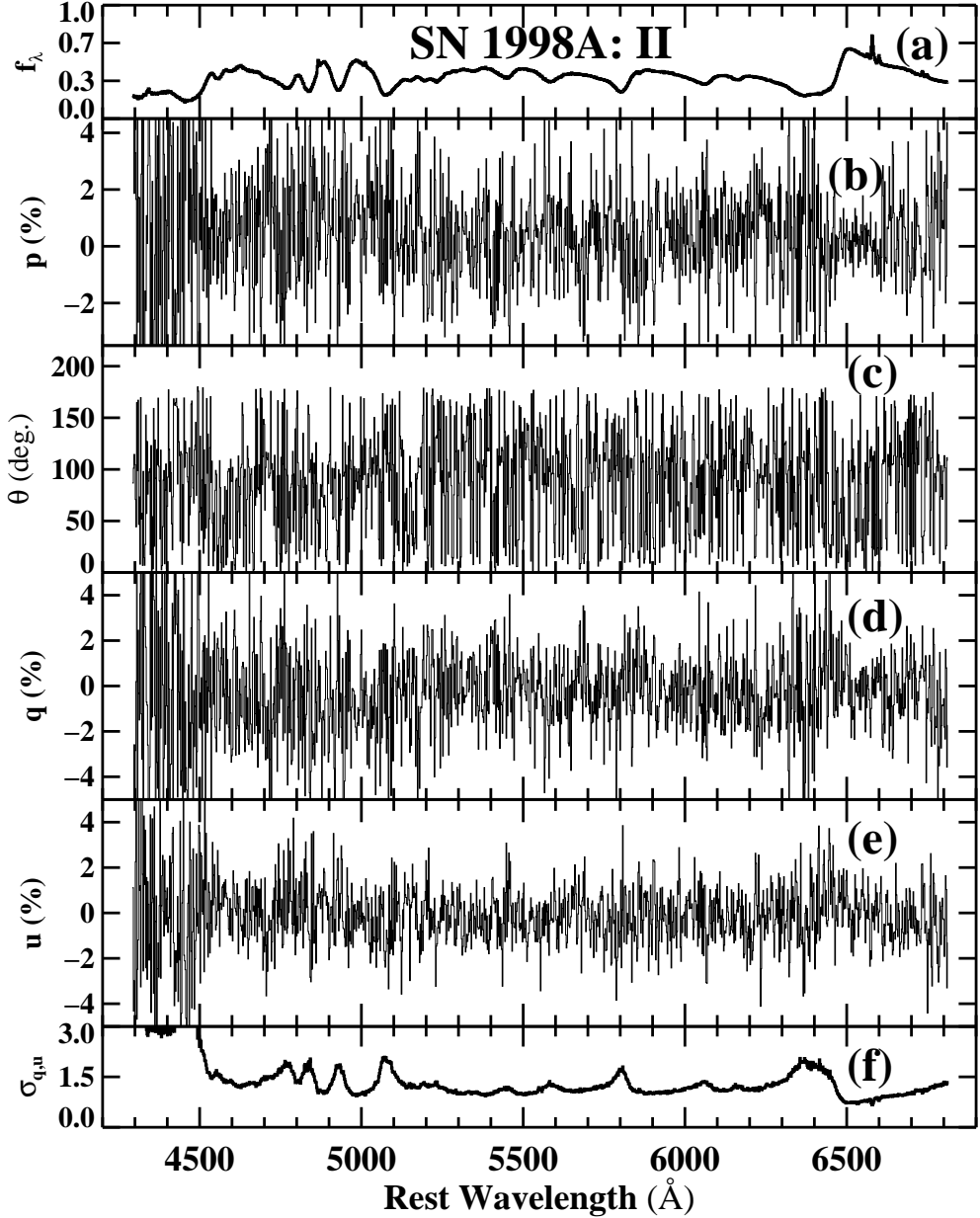}
		}
		}
\end{center}
\caption[Polarization data for SN~1998A]
{Polarization data for SN~1998A obtained 1998 January 17, between 11 and 30
days after explosion.  The NED recession velocity of 2081 km s$^{-1}$ for
IC~2627 has been removed in this and
all figures.  ({\it a}) Total flux, in units of $10^{-15}$ ergs
s$^{-1}$ cm$^{-2}$
\AA$^{-1}$.  ({\it b}) Observed degree of polarization.  ({\it c}) Polarization
angle in the plane of the sky. ({\it d, e}) The normalized $q$ and $u$ Stokes
parameters. ({\it f}) Average of the (nearly identical) $1\sigma$ statistical
uncertainties in the Stokes $q$ and $u$ parameters for the displayed binning of
2~\AA\ bin$^{-1}$.  
}
\label{fig4:1_98a}
\end{figure}

\begin{figure}
\ssp
\vskip -0.5in
\hskip -0.3in
%\begin{center}
\rotatebox{90}{
 \scalebox{0.8}{
	\plotone{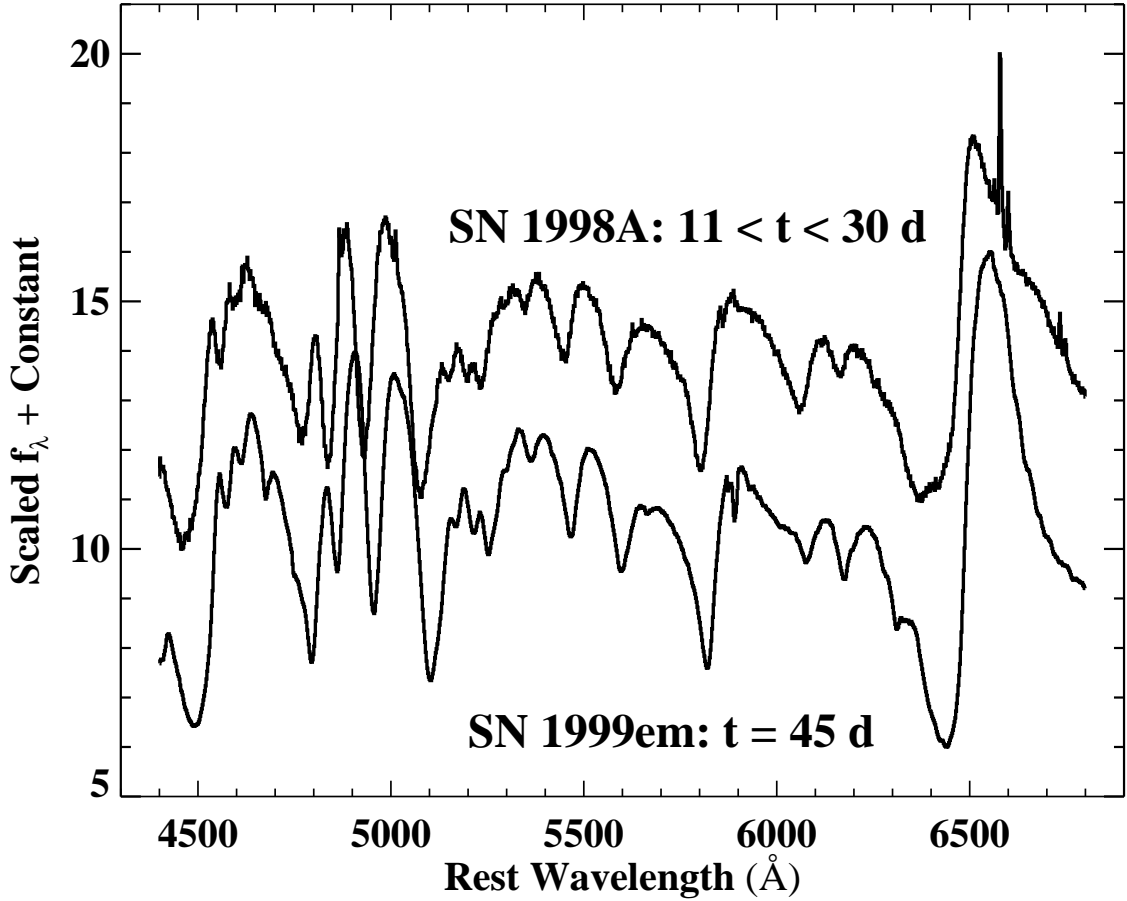}
		}
		}
%\end{center}
\caption[Total flux spectra of SN~1998A and SN~1999em compared]
{The total flux spectrum of SN~1998A ({\it top}) and SN~1999em ({\it bottom}),
with estimated time since explosion indicated.  The spectra have been offset
from each other for clarity.  Spectra of SN~1999em closer to the observational
epoch of SN~1998A have significantly weaker line features than SN~1998A.  The
more rapid spectral development of SN~1998A compared with SN~1999em is similar
to what was observed in spectra of SN~1987A.}
\label{fig4:5_98a}
\end{figure}

We obtained spectropolarimetry of SN~1998A on 1998 January 17 at the Keck-II
10-m telescope under cloudy conditions (Table~1).  From the constraints set
by the discovery and prediscovery images our observation likely took place
between $11$ and $30$ days after the explosion, when the SN was in its early
photospheric stage of development.  The data are presented in
Figure~\ref{fig4:1_98a} and tabulated in Table~2. The total flux spectrum
(Figure~\ref{fig4:1_98a}a) is characterized by prominent hydrogen Balmer and
P-Cygni features due to metal lines, typical for SNe~II during the
recombination phase.  The spectrum of SN~1999em that most closely matches the
line strengths of SN~1998A is from 45 days after explosion
(Figure~\ref{fig4:5_98a}); note that the strength of the metal lines in the
spectrum of SN~1999em from day 35 (Figure~\ref{fig4:5_97ds}) are considerably
weaker than those seen in the spectrum of SN~1998A from an even earlier epoch.
The more rapid spectral development of SN~1998A compared with SN~1999em is
consistent with what was seen in SN~1987A.  We also note that the depth of the
absorption feature near $6100$ \AA\ (generally attributed to a blend of
\ion{Ba}{2} $\lambda 6142$ along with several \ion{Fe}{2} features [L01b])
relative to the other metal lines is somewhat stronger in SN~1998A than in
SN~1999em.  It is tempting to speculate that this may be another spectral
similarity between SN~1998A and SN~1987A, since the spectrum of SN~1987A also
exhibited abnormally strong \ion{Ba}{2} features.  However, the contrast
between the depths of the \ion{Ba}{2} $\lambda 6142$ absorption and the other
metal lines was much greater in SN~1987A at a similar epoch than what is seen
here in SN~1998A.  Nevertheless, the available evidence does suggest that
spectral similarities between the two events do exist.

\begin{figure}
\ssp
\begin{center}
\rotatebox{0}{
 \scalebox{0.8}{ \plotone{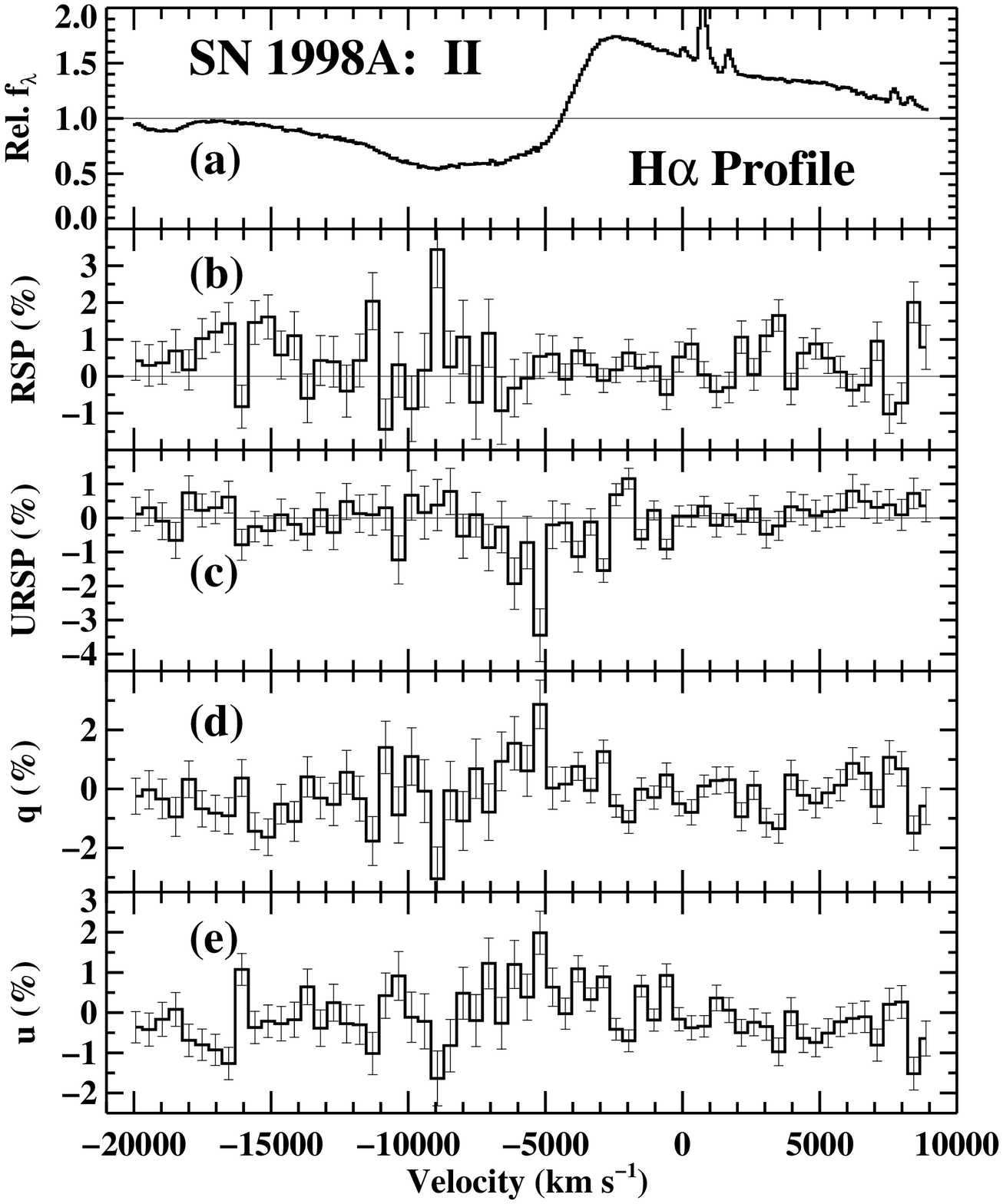}
	} }
\end{center}
\caption[Spectropolarimetry of H$\alpha$ region for SN~1998A] {Region
around H$\alpha$ for SN~1998A on 1998 January 17, between 11 and 30 days after
explosion.  Error bars are $1\sigma$ statistical for 10~\AA\ bin$^{-1}$.
({\it a}) Total flux, normalized by a spline fit to the continuum, displayed at
2~\AA\ bin$^{-1}$ for better resolution.  ({\it b}) Observed degree of
polarization, determined by the rotated Stokes parameter (rotated $q$).  ({\it
c}) Observed degree of polarization in the rotated $u$ parameter.  ({\it d, e})
Normalized $q$ and $u$ Stokes parameters.}
\label{fig4:2_98a}
\end{figure}

The optical polarimetry data shown in Figure~\ref{fig4:1_98a}(b,c) are quite
noisy, and we are unable to positively confirm the presence of any significant
line features, even at \halpha\ (Figure~\ref{fig4:2_98a}).  If we assume that
the width of a polarization feature at \halpha\ would be the same as that found
in SN~1997ds ($\sim 50$~\AA), then from the uncertainty in $q$ and $u$ we
derive upper limits of $\Delta q (3\sigma) = 1.6\%$ and $\Delta u (3\sigma) =
1.1\%$, so that $\Delta p_{tot} (3\sigma) = 1.9\%$, a detection limit somewhat
larger than the changes actually seen in SN~1997ds.  We therefore cannot place
very tight constraints on the degree of polarization change across line
features in SN~1998A.  The broadband continuum polarization, however, is quite
well defined, and found to be very low, $p_V = 0.24 \pm 0.05\%$ at $\theta =
118^{\circ} \pm 7^{\circ}$.  This polarization angle is inconsistent with the
polarization angles found for Galactic stars near the l-o-s, and suggests at
least some intrinsic SN polarization or host galaxy ISP contribution.  Since
${\rm ISP}_{max} = 0.9\%$, we conclude that $0\% < p < 1.14\%$ for SN~1998A,
and note that we cannot detect line features below $\Delta p_{tot} \approx 1.9
\%$.  The low level of polarization is similar to that observed in SN~1987A,
which exhibited $p_V \approx 0.2\%$ at early times followed by an increase to
$p_V \approx 0.7\%$ after about 40 days, and then a slow decline through day
100 (see Jeffery 1991a and references therein).

\subsection{SN~1999gi}
\label{sec:sn1999gi}

\begin{figure}
\ssp
\vskip -0.5in
\hskip -0.3in
%\begin{center}
\rotatebox{0}{
 \scalebox{1.0}{
	\plotone{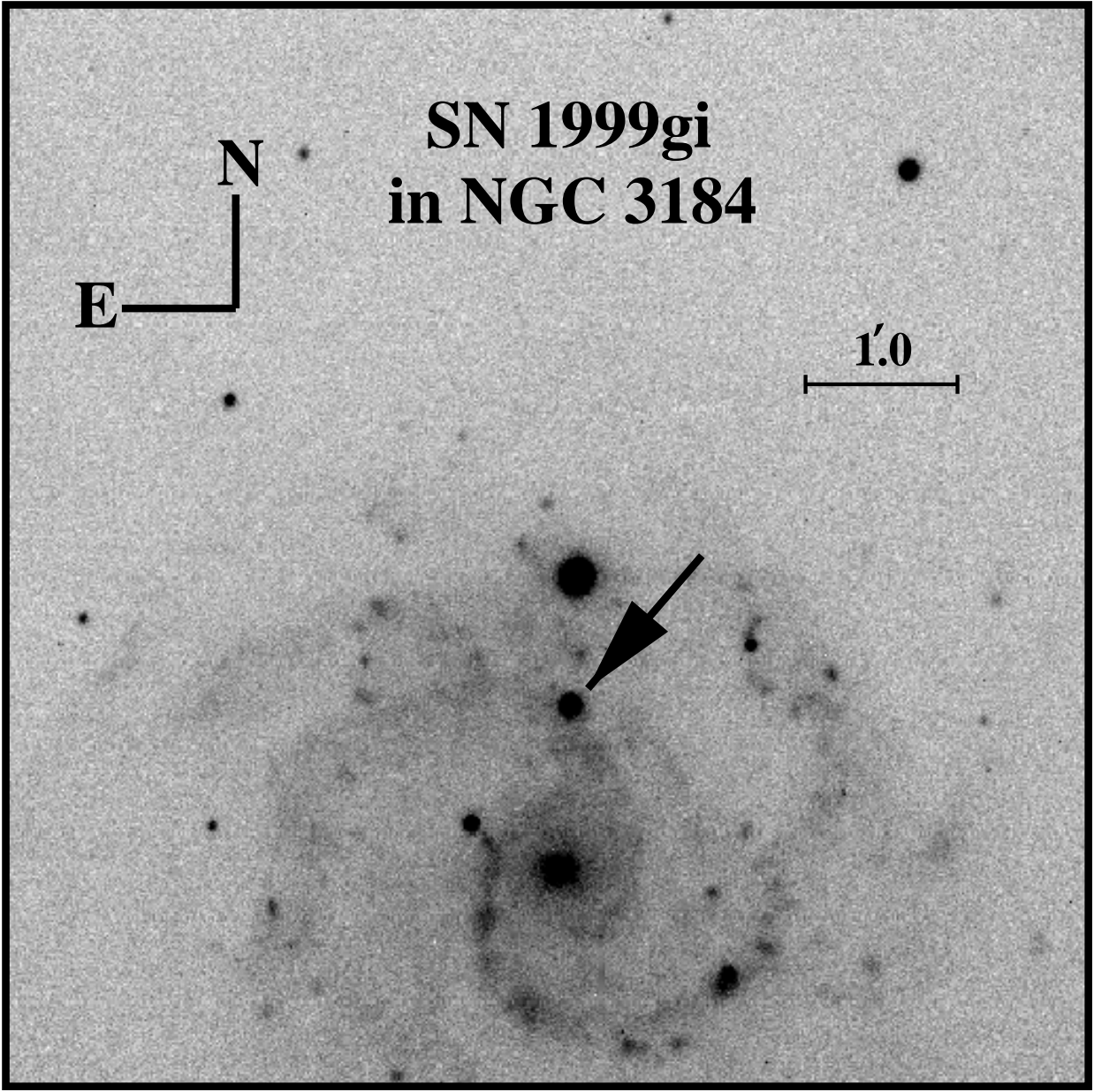}
		}
		}
%\end{center}
\caption[Image of SN~1999gi on 1999 December 17]
{$B$-band image of NGC 3184 taken on 1999 December 17 with the Katzman
Automatic Imaging Telescope (Treffers et al. 1997; Filippenko et al. 2001).
SN~1999gi ({\it arrow}) is measured to be 60.\arcsec9 north and 4.\arcsec7 west
of the center of NGC~3184 (cf., Nakano et al. 1999) .}
\label{fig4:7_99gi}
\end{figure}

SN~1999gi was discovered by Nakano et al. (1999) on 1999 December 9.82 at an
unfiltered magnitude of $m \approx 14.5$ mag in the nearly face-on ($i <
24^\circ$, from
LEDA\footnote{http://www-obs.univ-lyon1.fr/leda/home\_leda.html}) SBc galaxy
NGC~3184 (Figure~\ref{fig4:7_99gi}).  Since the SN is not visible on a CCD
image of the same field taken 7 days earlier\footnote{An additional
prediscovery image obtained 6 days prior to discovery also shows no SN (Trondal
et al. 1999), but it does not go as deep as the one taken the day before.}
(Nakano et al. 1999), it was likely discovered shortly after explosion.  A
spectrum taken within a day of discovery showed it to be a Type~II event at an
early epoch, with a blue continuum and a broad \halpha\ P-Cygni profile with a
hint of \ion{He}{1} $\lambda 5876$ absorption (Nakano et al. 1999).
Photometric data spanning the first 111 days after discovery clearly establish
SN~1999gi as a Type~II-P event, since its unfiltered magnitude fell by only
$\sim 0.4$ mag during this period (Kiss, Sarneczky, \& Sziladi 2000; Yoshida \&
Kadota 2000; Sarneczky et al. 2000).

The foreground reddening to SN~1999gi is minimal, \ebv $_{MW} = 0.02$ mag
(Schlegel et al. 1998).  However, there is significant \ion{Na}{1} D absorption
($W_\lambda {\rm [Na\ I\ D]} = 0.96$~\AA; Nakano et al. 1999) at the redshift
of NGC~3184, which suggests the possibility of substantial extinction by dust
in the host galaxy.  Applying equations~(\ref{eqn4:barbon}) and
(\ref{eqn4:pmax}) yields ${\rm ISP}_{max} = 2.34\% $, but we caution that
$W_\lambda = 0.96$~\AA\ is well into the saturation regime if few absorption systems
contribute to the line, so that the derived ISP$_{max}$ may underestimate the
true ISP contribution. A polarization measurement
to only one foreground MW star within $5^\circ$ of SN~1999gi is recorded in the
catalog by Heiles (2000), and it (HD 89021) has a very small value, $p = 0.01
\pm 0.12\%$, but is only 30 pc away and therefore does not nearly probe all of the
material in the Galactic plane (a star would need to be more than 183 pc away
since SN~1999gi has a Galactic latitude of $55^\circ$).  Nonetheless, from the
small Galactic reddening we expect a small ISP$_{MW}$.
\begin{figure}
\ssp
\begin{center}
\rotatebox{0}{
 \scalebox{0.8}{
	\plotone{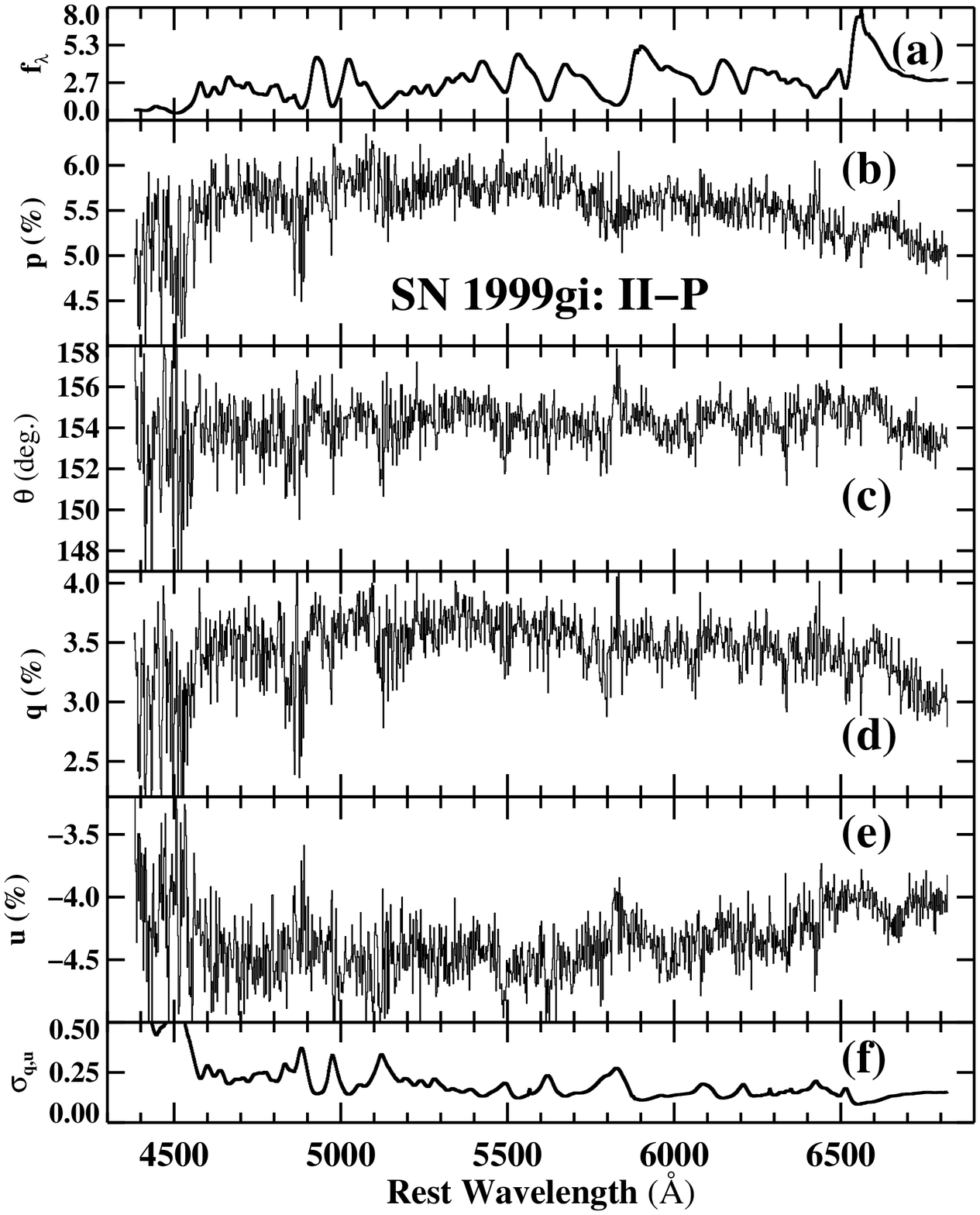}
		}
		}
\end{center}
\caption[Polarization data for SN~1999gi]
{Polarization data for SN~1999gi obtained on 2000 March 25, between 107 and 114
days after explosion.  The NED recession velocity of 532 km s$^{-1}$ for
NGC~3184 has been removed in this and
all figures.  ({\it a}) Total flux, in units of $10^{-15}$ ergs
s$^{-1}$ cm$^{-2}$~\AA$^{-1}$.  ({\it b}) Observed degree of polarization.
({\it c}) Polarization angle in the plane of the sky. ({\it d, e}) The
normalized $q$ and $u$ Stokes parameters. ({\it f}) Average of the (nearly
identical) $1\sigma$ statistical uncertainties in the Stokes $q$ and $u$
parameters for the displayed binning of 2~\AA\ bin$^{-1}$.  }
\label{fig4:1_99gi}
\end{figure}

\begin{figure}
\ssp
\begin{center}
\rotatebox{0}{
 \scalebox{0.8}{ \plotone{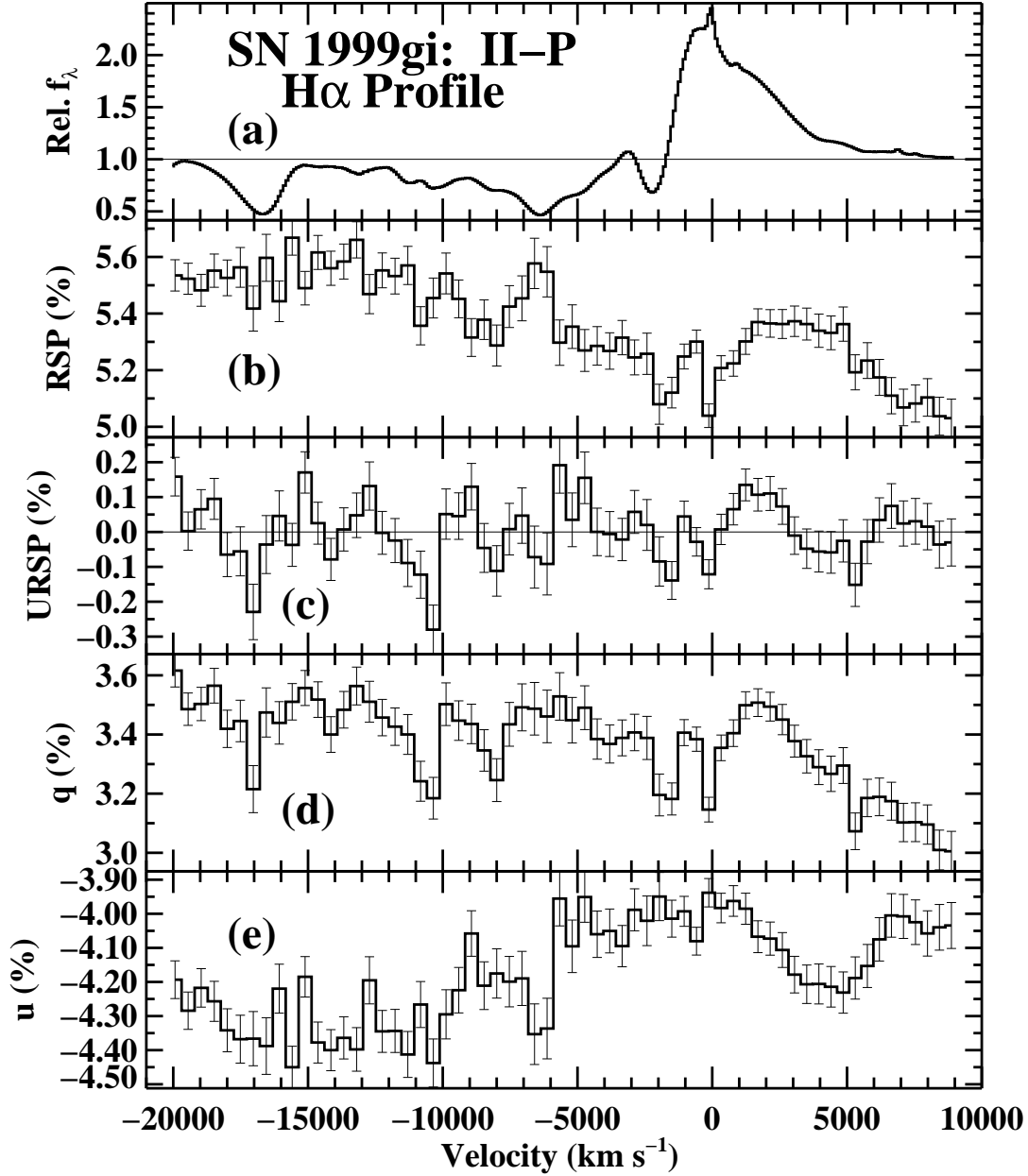}
	} }
\end{center}
\caption[Spectropolarimetry of H$\alpha$ region for SN~1999gi] {Region
around H$\alpha$ for SN~1999gi on 2000 March 25, between 107 and 114 days after
explosion.  Error bars are $1\sigma$ statistical for 10~\AA\ bin$^{-1}$.  ({\it
a}) Total flux, normalized by a spline fit to the continuum, displayed at
2~\AA\ bin$^{-1}$ for better resolution.  ({\it b}) Observed degree of
polarization, determined by the rotated Stokes parameter (rotated $q$).  ({\it
c}) Observed degree of polarization in the rotated $u$ parameter.  ({\it d, e})
Normalized $q$ and $u$ Stokes parameters.}
\label{fig4:2_99gi}
\end{figure}

We obtained high-quality spectropolarimetry of SN~1999gi on 2000 March 25 (107
days after discovery) at the Keck-I 10-m telescope (Table~1).  From the
constraints set by the discovery and prediscovery images our observation likely
took place between $107$ and $114$ days after the explosion.  A photometric
observation by Sarneczky et al. (2000) 4 days after our observation showed
SN~1999gi to be at unfiltered magnitude $m = 14.9$ mag; since SN~1999gi
reportedly dropped an additional 0.3 mag over the next 11 days (Sarneczky et
al. 2000), it is likely that SN~1999gi was right at the end of the plateau
phase when our observation took place.  The observed spectropolarimetry data
are shown in Figure~\ref{fig4:1_99gi}, with detail of the \halpha\ region
displayed in Figure~\ref{fig4:2_99gi}.

\begin{figure}
\ssp
\vskip -0.5in
\hskip -0.3in
%\begin{center}
\rotatebox{90}{
 \scalebox{0.8}{
	\plotone{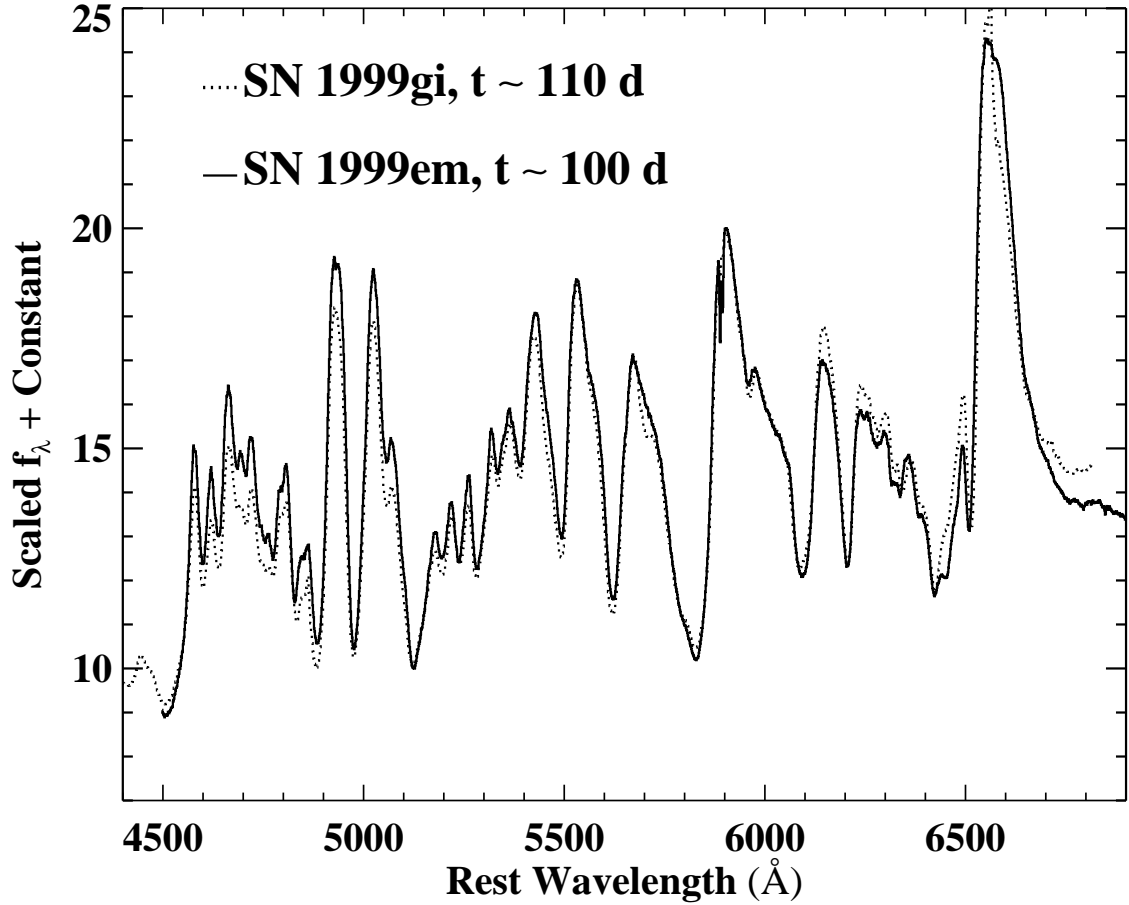}
		}
		}
%\end{center}
\caption[Total flux spectra of SN~1999gi and SN~1999em compared]
{The total flux spectrum of SN~1999gi taken between 107 and 114 days after
explosion ({\it dotted line}) compared with a spectrum of SN~1999em roughly 100
days after explosion ({\it solid line}), with the spectra overplotted to
demonstrate the extreme similarity of the line profiles.  The similarity of all
the features, particularly the complex \halpha\ absorption profile, suggests a
similar evolutionary phase for the two events.}
\label{fig4:5b_99gi}
\end{figure}

Figure~\ref{fig4:5b_99gi} illustrates the striking similarity between the total
flux spectrum of SN~1999gi and a spectrum of SN~1999em taken at a similar epoch
($\sim 100$ days after explosion).  Of particular interest is the remarkable
correspondence of all of the absorption features that complicate the \halpha\
absorption profile.  Since rapid changes in the \halpha\ profile of SN~1999em
coincided with its fall off the plateau (L01b), this strengthens the argument
that SN~1999gi was observed at a similar evolutionary stage in its development.
A detailed discussion of the spectral development of SNe~II-P, with a
particular focus on the \halpha\ profile, is given by L01b.  For this study, we
note that the existing spectroscopic and photometric evidence suggests that
SN~1999gi and SN~1999em may have been quite similar events and that our
spectropolarimetric observation of SN~1999gi likely caught the evolutionary
phase when the recombination front was just reaching the H$-$He boundary and
the SN was beginning the transition to the nebular stage.

\begin{figure}
\ssp
\vskip -0.5in
\hskip -0.3in
%\begin{center}
\rotatebox{90}{
 \scalebox{0.8}{
	\plotone{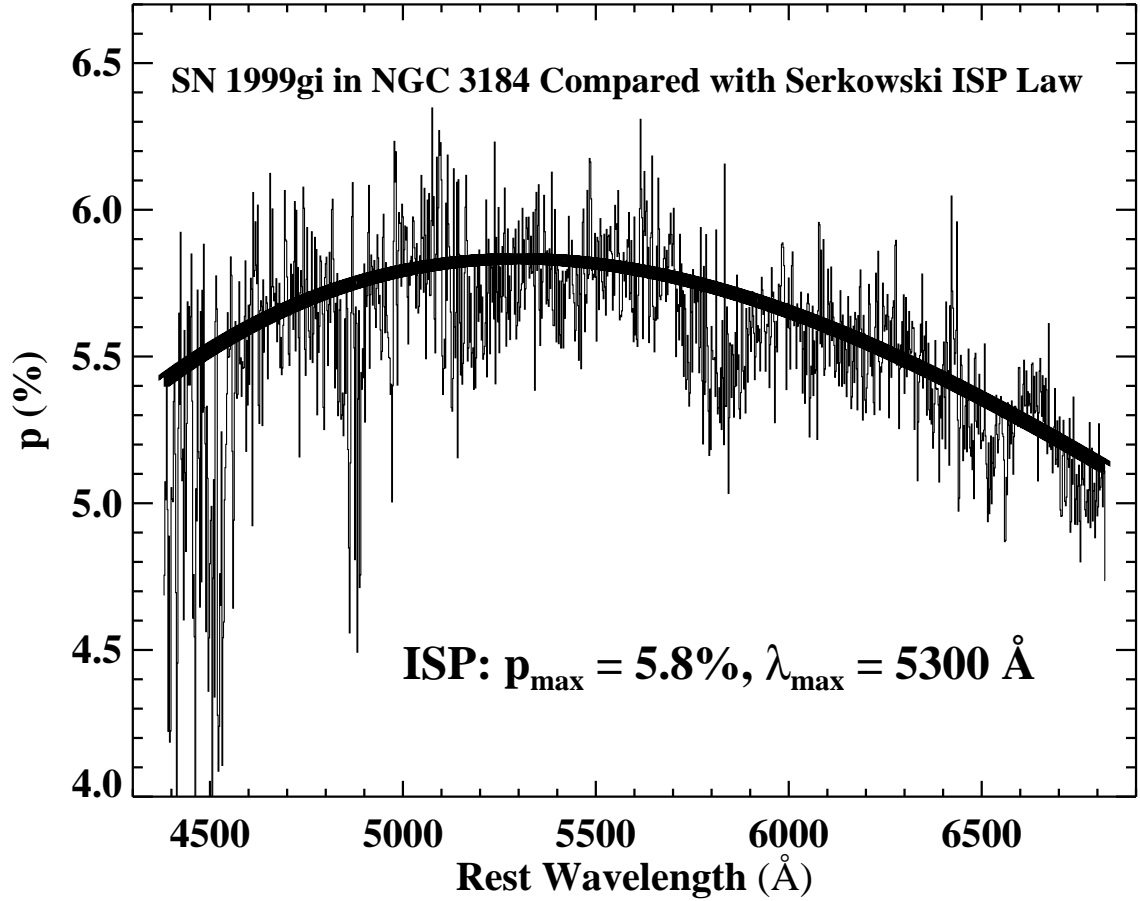}
		}
		}
%\end{center}
\caption[Observed polarization of SN~1999gi compared with a Serkowski ISP
curve] {Observed polarization of SN~1999gi ({\it thin line}) compared with the
best fitting Serkowski ISP curve ({\it thick line}) characterized by $p_{\rm
max} = 5.8\%, \theta = 154^\circ, {\rm and\ } \lambda_{\rm max} = 5300$~\AA. }
\label{fig4:3_99gi}
\end{figure}

\begin{figure}
\ssp
\vskip -0.5in
\hskip -0.3in
%\begin{center}
\rotatebox{90}{
 \scalebox{0.8}{
	\plotone{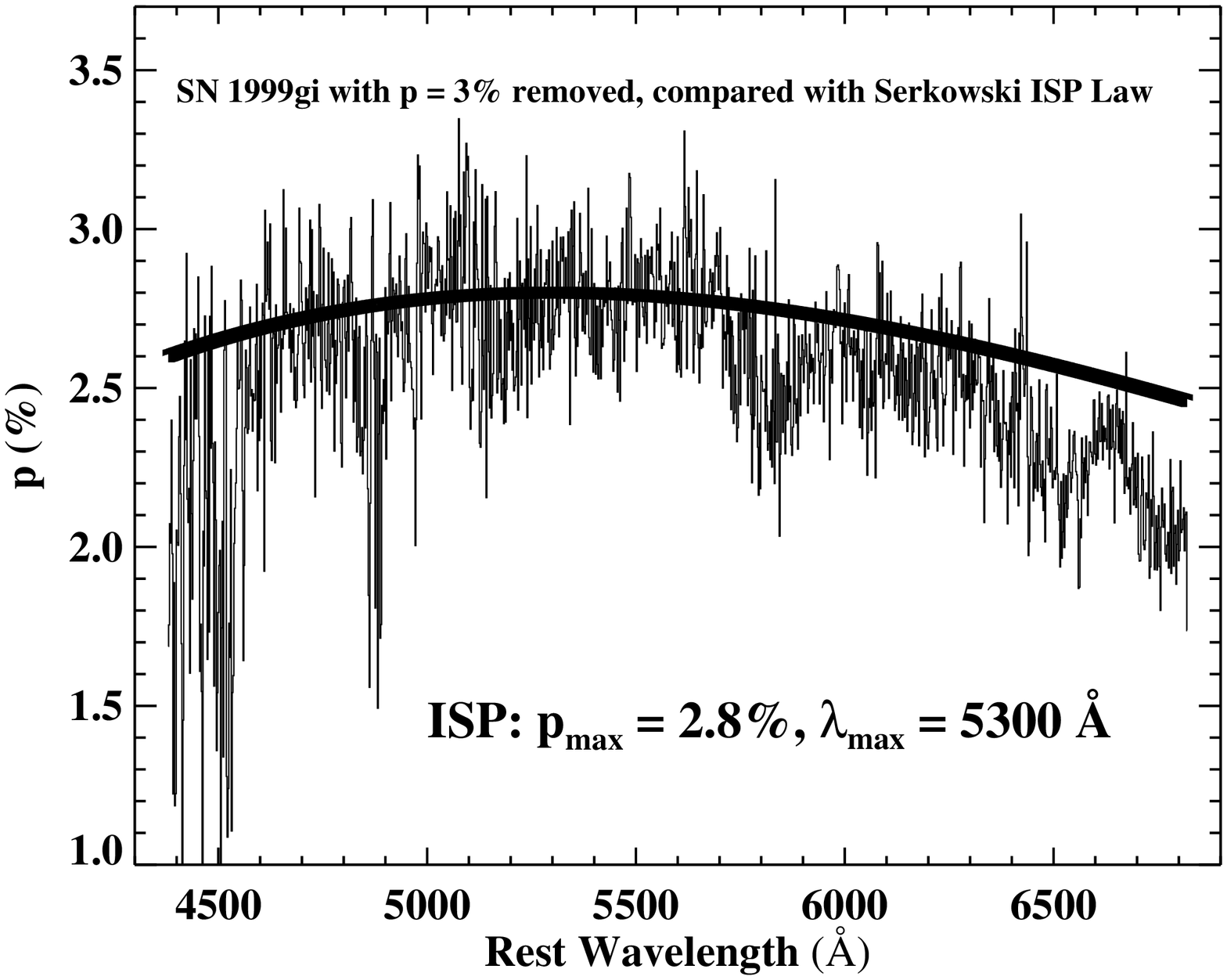}
		}
		}
%\end{center}
\caption[Polarization of SN~1999gi with arbitrary SN polarization removed,
compared to a Serkowski ISP curve] {Observed polarization of SN~1999gi with a
wavelength-independent polarization of $p = 3\%$, $\theta = 154^\circ$ removed
({\it thin line}), compared with the best-fitting Serkowski ISP curve ({\it
thick line}), characterized by $p_{\rm max} = 2.8\%, \theta = 154^\circ, {\rm
and\ } \lambda_{\rm max} = 5300$~\AA. The continuum polarization fit here is
not as convincing as in Figure~\ref{fig4:3_99gi}, suggesting that ISP due to
dust in NGC~3184 may dominate the polarization signal.}
\label{fig4:3a_99gi}
\end{figure}

SN~1999gi exhibits an extremely high overall polarization level, $p_V = 5.72
\pm 0.01\%$ (Table~2), with polarization modulations evident across several
P-Cygni features.  It is unfortunate that no other spectropolarimetry (or
broadband polarimetry) exists for SN~1999gi, since a temporal change in the
overall level would help distinguish between polarization intrinsic to the SN
and that contributed by ISP.  The main question, then, is what fraction of the
observed polarization is due to ISP, and how much is intrinsic to SN~1999gi.
With only a single epoch of spectropolarimetry, it is impossible to know for
sure.  However, there is one characteristic of its continuum polarization that
sets it apart from all other Type II objects studied thus far: it is not flat
with wavelength.  Rather, it has a broad, asymmetric peak at about $5300 \pm
100$~\AA\ that gently decreases on either side.  This is unlike the
wavelength-independent nature of electron scattering expected for a SN
atmosphere, but is reminiscent of the polarization seen in spectropolarimetry
of Galactic stars whose light is polarized by foreground dust.  In fact,
Figure~\ref{fig4:3_99gi} demonstrates that a Serkowski ISP curve (Serkowski
1973) as modified by Wilking, Lebofsky, \& Rieke (1982) and updated by Whittet
et al. (1992), convincingly fits the continuum polarization exhibited by
SN~1999gi.  When arbitrary amounts of wavelength-independent polarization are
removed from the observed polarization (to simulate polarization intrinsic to
SN~1999gi), the fit of the best-fitting Serkowski ISP curve becomes less
convincing (Figure~\ref{fig4:3a_99gi}).  Since the fit of a Serkowski ISP curve
to the resulting ISP becomes quite poor for intrinsic SN polarization values
greater than $\sim 2\%$, this implies that the polarization intrinsic to
SN~1999gi is somewhat less than $2\%$, and that ISP$_{host}$ is likely greater
than $\sim 3.7\%$.  This suggests (but does not prove) that much of the
polarization seen is, in fact, interstellar in nature.  We shall return to a
discussion of the potential power of SN spectropolarimetry for the study of the
properties of dust in external galaxies in \S~\ref{sec:dustprobe}.  For now,
we note that the shape, strength, and peak wavelength of the polarization seen
in SN~1999gi is quite consistent with heavily reddened lines of sight to
Galactic stars, which would imply similar properties for the dust in NGC~3184
and the MW.

Along these same lines, the constancy of the polarization angle with wavelength
across the optical spectrum (Figure~\ref{fig4:1_99gi}c) also favors a large
ISP$_{host}$. If large intrinsic SN~polarization exists, then unless
$\theta_{\rm SN} = \theta_{\rm ISP}$ (an unlikely chance occurrence),
significant variation in the polarization angle with wavelength would be
observed.  For the same reason, this also favors the idea that ISP$_{host}$
comes predominantly from a single source, perhaps a thick dust cloud along the
l-o-s, in which much of the gas has been depleted onto dust grains.  This
scenario would also explain how ISP$_{host}$ could be so much greater than the
ISP$_{max}$ predicted by equation~(\ref{eqn4:pmax}).

\begin{figure}
\ssp
\vskip -0.5in
\hskip 0.3in
%\begin{center}
\rotatebox{90}{
 \scalebox{0.8}{
	\plotone{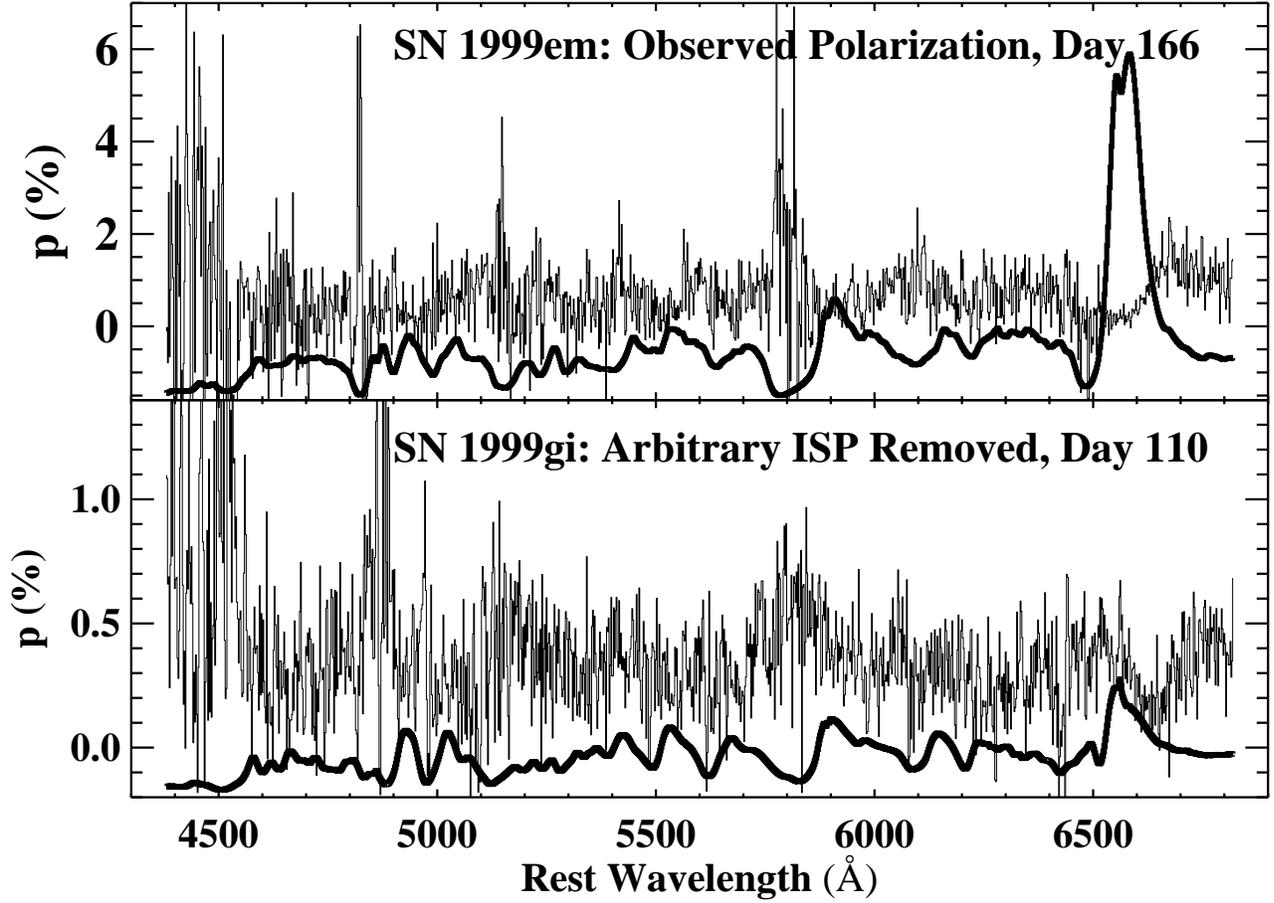}
		}
		}
%\end{center}
\caption[``Intrinsic'' polarization of SN~1999gi, after arbitrary ISP removed]
{A comparison of the polarization ({\it thin lines}) and total flux ({\it thick
lines}) properties of SN~1999em and SN~1999gi, observed at somewhat different
epochs.  The polarization of SN~1999gi is shown after removal of an arbitrary
ISP characterized by a Serkowski ISP curve with $p_{max} = 6.1\%, \theta =
154^\circ, {\rm and\ } \lambda_{max} = 5300$~\AA; this ISP was chosen to
demonstrate the possible similarity between SN~1999gi and SN~1999em, and is not
a unique solution. }
\label{fig4:4_99gi}
\end{figure}

A final argument in favor of a large ISP$_{host}$ contribution comes from the
low contrast of the polarization modulations seen across the P-Cygni lines.
The two strongest polarization features, near $\lambda = 4500 {\rm\ \AA\ }$
(\ion{Ba}{2} $\lambda 4554\ +\ {\rm Fe\ II\ blend}$) and $\lambda = 5800 {\rm\
\AA\ }$ (\ion{Na}{1} D), have $\Delta p_{tot} \approx 1.0\%$ and $\Delta
p_{tot} \approx 0.6\%$, respectively.  This level of polarization change is
quite similar to that observed in the line troughs of SN~1997ds and SN~1999em.
If the line-trough polarization mechanism in SN~1999gi is the same as that
proposed for SN~1999em and SN~1997ds (\S~\ref{sec:modulations}), much stronger
features should exist in SN~1999gi if it has a more highly polarized continuum.
In fact, since the \ion{Ba}{2} $\lambda 4554$ blend has $I_{cont} / I_{trough}
\approx 3$ and the \ion{Na}{1} D line has $I_{cont} / I_{trough} \approx 4$,
this sets lower bounds on the continuum polarization of only $p_{cont} >
0.25\%$ and $p_{cont} > 0.30\%$, respectively (equation~[\ref{eqn4:pcont}]),
for SN~1999gi.  A further similarity between SN~1999gi and SN~1999em is the
polarization modulation seen across the \halpha\ profile, especially the
changes in the red emission wing of the profile (see L01a, Figs. 14, 15, and
16).  Purely to highlight the possible similarity between SN~1999gi and
SN~1999em, we show in Figure~\ref{fig4:4_99gi} the ``intrinsic'' SN
polarization that results for SN~1999gi after removing an arbitrary (large) ISP
compared with the observed polarization of SN~1999em (which is though to suffer
minimal ISP) from a later epoch.  While this comparison does not prove an
intrinsic similarity between the polarization of SN~1999gi and SN~1999em since
the ISP is not known (and different ISP values can even make polarization
maxima become minima and vice versa), it does show that ISP values exist that
do make the spectropolarimetry of SN~1999gi resemble that observed in
SN~1999em.  Regardless of the ISP, though, the strength of the line features in
SN~1999gi is not significantly different from that observed in either SN~1999em
or SN~1997ds.

Although the evidence for large ISP$_{host}$ is compelling, we must consider
arguments for significant intrinsic polarization for SN~1999gi as well.  In
fact, there are both theoretical and observational reasons supporting a large
intrinsic SN polarization.  We first present a theoretical argument.  The low
polarization observed in SN~1999em at early times coupled with the relatively
low polarization constraints set for SN~1998A and SN~1997ds leads to the
tentative conclusion that SNe~II with significant hydrogen envelopes intact at
the time of explosion are substantially spherical at early times.  However,
there is mounting evidence supporting the idea of strongly asymmetric
explosions, including the temporal increase in polarization observed in
SN~1999em.  The observations of SN~1999em, coming on days 12, 45, 54, and 166
after explosion, as well as the observation of SN~1997ds $33 - 46$ days after
explosion and SN~1998A $11 - 30$ days after explosion, all missed the crucial
evolutionary phase at the end of the plateau, when the hydrogen envelope has
nearly completely recombined, but before the SN begins the transition to the
nebular phase.  At this critical juncture the photosphere has receded through
nearly the entire hydrogen envelope, yet a thermal photosphere with a large
electron-scattering optical depth still exists.  This creates an ideal
situation for explosion asymmetry to be revealed by spectropolarimetry: if any
epoch is to reveal the effects of great explosion asymmetry, this would seem to
be it.  At much earlier times the photosphere is located in the extended
hydrogen envelope, and at much later times (i.e., the day 166 observation of
SN~1999em), the electron-scattering optical depth may be quite small.
Therefore, the timing of the observation of SN~1999gi is unique among the
objects thus far studied, and it may not be fair to favor low intrinsic
polarization based on comparisons with other SNe~II observed at substantially
earlier or later epochs.

A simple observational argument favoring significant intrinsic SN polarization
follows from the few optical imaging polarimetric studies that have been done
of nearly face-on spiral galaxies (e.g., Scarrott, Rolph, \& Semple 1990;
Scarrott et al. 1991).  These studies generally find polarization vectors
aligned parallel to the spiral arms (i.e., perpendicular to the line connecting
a point with the galaxy's center).  From Figure~\ref{fig4:7_99gi}, we see that
SN~1999gi is nearly due north of the center of NGC~3184, which would predict
$\theta_{ISP} \approx 90^{\circ}$, quite different from the measured value of
$\theta = 154^{\circ}$.  However, since the number of galaxies studied is very
small and, unlike the l-o-s to a SN, optical imaging polarimetry does not probe
very deeply into dusty regions (see, e.g., Jones 1997), the inconsistency
between the observed polarization angle and that predicted may not be too
significant.  Also, the lack of appreciable spectral polarization angle
variation (Figure~\ref{fig4:1_99gi}) argues for a single, dominant,
polarization source.

Given the preceding arguments and lacking additional polarimetric epochs or
independent estimates of the ISP, it is quite difficult to assign a value to
ISP$_{host}$ with absolute confidence.  On the whole, the low contrast of the
line features, the excellent agreement between a Serkowski ISP curve and the
observed continuum polarization shape, and the near constancy of the polarization
angle with wavelength lead us to favor the interpretation that at least a {\it
majority} of the observed polarization is due to ISP$_{host}$.  We allow for
the possibility, however, that SN~1999gi may still have significantly more
intrinsic polarization than the other SNe II studied.  From the limits
prescribed by the strength of the line features and the goodness of the
Serkowski ISP curve fit, our preferred interpretation of the data is that
SN~1999gi is intrinsically polarized at $0.30\% < p < 2.0\%$.  Low-contrast 
modulations (up to $\Delta p_{tot} = 1.0\%$) in the troughs of the strongest
P-Cygni lines and across the entire \halpha\ emission profile are also
observed.

\subsection{SN Spectropolarimetry as a Probe of Interstellar Dust in External
Galaxies}
\label{sec:dustprobe}

Quantifying the reddening properties of interstellar dust in all but the
closest external galaxies is difficult (see, e.g., Fitzpatrik 1998).  The good
agreement between the polarization of SN~1999gi and a Serkowski ISP curve
(Figure~\ref{fig4:3_99gi}) highlights a potentially powerful byproduct of
SN~spectropolarimetry: SNe can serve as probes of the interstellar medium of
external galaxies, providing information about the alignment and size
distribution of the galaxy's dust grains, its reddening law, and its magnetic
field strength and direction (see, e.g., Spitzer 1978).

Polarimetric studies of reddened Galactic stars whose light is intrinsically
unpolarized have established a convincing correlation between the wavelength of
maximum polarization ($\lambda_{max}$) and the ratio of total to selective
extinction (Whittet \& van Breda 1978):
\begin{equation}
R_V = (5.6 \pm 0.3) \lambda_{max} {\rm (\AA)} / 10000,
\label{eqn4:rv}
\end{equation}
\noindent where  $R_V \equiv A_V / \Ebv$. Values of $\lambda_{max}$ generally
fall in the range $4000 - 8000$~\AA, with a mean value of $5500$~\AA\ providing
the canonical $R_V = 3.1$ (e.g., Savage \& Mathis 1979).  If we assume that
most of the polarization of SN~1999gi is due to ISP$_{host}$, then our
measured $\lambda_{max} = 5300 \pm 100$~\AA\ yields $R_V = 3.0 \pm 0.2$ for the dust along this
l-o-s in NGC~3184\footnote{We note that due to the vector nature of
polarization, intrinsic SN polarization, even if independent of wavelength,
could conspire with the ISP to produce a $\lambda_{max}$ different from that
produced by the ISP alone.  Given the constancy of its polarization angle
(\S~\ref{sec:sn1999gi}; Figure~\ref{fig4:1_99gi}c), however, this seems
unlikely for SN~1999gi. }.  If the reddening to SN~1999gi could be determined,
then the level of ISP$_{host}$ relative to \ebv\ would give a measure of the
efficiency of grain alignment and hence the strength of the magnetic field in
NGC~3184 (e.g., Johnson 1982).

The value of $\lambda_{max}$ is thought to be related to the average size of
the polarizing dust grains through the relation (Whittet 1992)
\begin{equation}
\lambda_{max} \approx 2\pi a(n - 1),
\label{eqn4:size}
\end{equation}
\noindent where $a$ is the mean characteristic size of the dust grains and $n$
is the index of refraction, with $n = 1.6$ being typical for silicates.  A
$\lambda_{max}$ of $5300 \pm 100$~\AA\ therefore suggests dust grains in
NGC~3184 along the l-o-s to SN~1999gi that have an average size of $\sim 0.14\
\micron$, similar to the inferred size of typical dust grains in the MW.

We conclude that the interstellar dust polarizing the light of SN~1999gi has a
similar size distribution and reddening characteristics as average MW dust.
This result is consistent with the results of the study by Riess, Press, \&
Kirshner (1996), in which the {\it average} $R_V$ value in galaxies hosting SNe
Ia is shown to be consistent with a typical Galactic extinction law.

Spectropolarimetry of SNe can therefore be a powerful diagnostic tool in the
study of the ISM of external galaxies.  Of course, the possibility of intrinsic
SN~polarization complicates their use as probes.  As more individual SNe
continue to be studied with spectropolarimetry, however, the intrinsic
polarization characterizing each SN class should become better defined.  The
best probes will be the types of SNe found to have consistently low intrinsic
polarization.  Although too few events have been studied to warrant firm
generalizations, existing evidence suggests that SNe~Ia (e.g., Hough et
al. 1987; Wang, Wheeler, \& H\"{o}flich 1997), and perhaps SNe~II-P at early
stages, may be the most promising candidates.

\section{Conclusions}
\label{sec:conclusions}

We present single-epoch optical spectropolarimetry of three SNe~II during the
photospheric phase.  None of the objects shows evidence for significant
circumstellar interaction, and all probably had most of their hydrogen envelope
intact at the time of explosion.  Our main results are as follows.

\begin{enumerate}
\item SN~1997ds is a Type~II event that bears spectral similarities to the Type
II-P SN~1999em at a similar observational epoch.  Lacking photometric data,
however, we cannot assign it a definitive subclassification (i.e., II-P or
II-L).  Our spectropolarimetric observation of SN~1997ds took place between 33
and 46 days after the explosion, and we measure $p_V = 0.85 \pm 0.02\%$, with a
distinct polarization modulation of $\Delta p_{tot} \approx 1.6\% $ in the
\halpha\ absorption trough.  Modulations greater than $1\%$ in other absorption
features are not seen.  From reddening considerations we conclude that
SN~1997ds is intrinsically polarized by $0.04\% \leq p \leq 1.66\%$.

\item SN~1998A is a Type~II event, sharing some photometric and spectroscopic
properties with the unusual core-collapse event, SN~1987A.  Our
spectropolarimetry data were obtained between 11 and 30 days after the
explosion, and we measure $p_V = 0.24 \pm 0.05\%$.  From the low estimated
degree of reddening, we conclude that the intrinsic polarization of SN~1998A
is $0\% < p < 1.14\%$.  Although we find no evidence for polarization
modulations across strong line features, we note that our data are only
sensitive enough to detect sharp features above $\Delta p_{tot} \approx 1.9
\%$.

\item SN~1999gi is a Type~II-P event, with a plateau that lasted at least until
our spectropolarimetric observation took place, between 107 and 114 days after
the explosion.  The total flux spectrum of SN~1999gi is nearly identical to a
spectrum of SN~1999em obtained at a similar epoch.  SN~1999gi has a very high
observed polarization, $p = 5.72\%$, with distinct changes of up to $\Delta
p_{tot} = 1.0\%$ seen in the troughs of the strongest P-Cygni lines.  A strong
polarization change across the entire \halpha\ profile is also seen.  Since the
spectral shape of the continuum polarization closely resembles a Serkowski ISP
curve and is inconsistent with the wavelength-independent nature of electron
scattering, we believe the majority of the observed polarization to be due to
ISP from dust in the host galaxy, NGC~3184. Significant
intrinsic SN polarization cannot be ruled out, however.  Our preferred
interpretation is that SN~1999gi is intrinsically polarized between $0.3\%$ and
$2.0\%$.

\end{enumerate}

Although different in detail, SNe~1997ds, 1998A, and 1999gi all share
common spectropolarimetric properties with SN~1999em.  Most importantly,
neither SN~1997ds nor SN~1998A show compelling evidence for large intrinsic
polarization at the early phases studied.  This furthers the argument that the
electron-scattering atmospheres of SNe~II with significant hydrogen envelopes
may in general suffer from only modest asphericity during the early
recombination phase.  This is in contrast with what has been observed core-collapse
events involving progenitors that have lost significant amounts of their
envelopes prior to exploding (e.g., Tran et al. 1997; Wang et al. 2001), or are
interacting with a dense CSM (Leonard et al. 2000a).

The number of core-collapse events studied in detail with spectropolarimetry is
still quite small.  However, the basic result that evidence for asphericity
increases the deeper one probes into the heart of the explosion is consistent
with this study.  The possibility that SN~1999gi may have possessed a larger
intrinsic polarization near the end of the plateau phase than the other events
studied at earlier phases would also support this trend.  These results
therefore support the hypothesis advanced by L01a that distances derived to
SNe~II-P through the expanding photosphere method at early photospheric epochs
do not suffer large uncertainties resulting from asphericity.

\bigskip
\smallskip

We thank Aaron Barth, Edward Moran, and Maryam Modjaz for assistance with the
observations, and Tom Matheson and Weidong Li for useful discussions.  Most of
the data presented herein were obtained at the W. M. Keck Observatory, which is
operated as a scientific partnership among the California Institute of
Technology, the University of California, and the National Aeronautics and
Space Administration.  The Observatory was made possible by the generous
financial support of the W. M. Keck Foundation. We are grateful to the Keck
staff for their support of the telescopes.  This research has made use of the
NASA/IPAC Extragalactic Database (NED), which is operated by the Jet Propulsion
Laboratory, California Institute of Technology, under contract with NASA.  We
have made use of the LEDA database (http://leda.univ-lyon1.fr).  Our work was
funded by NASA grants GO-7821, GO-8243, and GO-8648 from the Space Telescope
Science Institute, which is operated by AURA, Inc., under NASA contract NAS
5-26555.  Additional funding was provided to A. V. F. by NASA/Chandra grant
GO-0-1009C, by NSF grants AST-9417213 and AST-9987438, by the Sylvia and Jim
Katzman Foundation, and by the Guggenheim Foundation.

\end{document}